\documentclass[sigplan,nonacm]{acmart}

\usepackage{tikz}
\usepackage{cleveref}
\usepackage{amsmath}

\usepackage{algorithmic}
\usepackage{graphicx}
\usepackage{textcomp}
\usepackage{xcolor}

\definecolor{indigo}{RGB}{51,0,102}
\usepackage{tcolorbox}

\usepackage{hyperref}

\usepackage[framemethod=TikZ]{mdframed}

\usepackage{xspace}
\usepackage{scalerel}
\usepackage{ctable}

\usepackage{pifont}

\usepackage{url}
\usepackage{mathtools}

\usepackage{tabularray}

\usepackage{multirow}
\usepackage{array}
\usepackage[normalem]{ulem}

\AtBeginDocument{%
  }

\usepackage{listings}
\usepackage{algorithmic}
\usepackage{graphicx}
\usepackage{textcomp}
\usepackage{xcolor}
\usepackage[user,hyperref]{zref}
\usepackage{tablefootnote}
\usepackage{xspace}
\usepackage{algorithmic}
\usepackage{textcomp}
\usepackage{multirow}
\usepackage{xcolor}
\usepackage{circledsteps}
\usepackage[ruled,linesnumbered]{algorithm2e}
\usepackage{soul}
\usepackage{xurl}

\usepackage{booktabs} 
\usepackage{multirow} 
\usepackage{siunitx}

\newcommand\mypagebreak{\clearpage\newpage}
\renewcommand\mypagebreak{}

\definecolor{brightpink}{rgb}{1.0, 0.0, 0.5}

\newcommand{\sysname}{Penrose\xspace}
\DeclareRobustCommand{\takeaway}[1]{}%
\AtBeginEnvironment{quote}{
\it\small}
\newcounter{mynote}[section]

\newcommand{\CS}[0]{AS}

\newcommand{\func}{\textsf}

\newcommand{\secref}[1]{\mbox{\S~\ref{#1}}}

\newcommand{\mus}{\text{\textmu s}}  

\begin{document}
\title{Penrose: Privacy-Preserving Performance Profiling of In-The-Wild GPUs}

\author{Ian McDougall}
\email{imcdougall@wisc.edu}
\affiliation{%
  \institution{University of Wisconsin-Madison}
  \city{Madison}
  \state{WI}
  \country{USA}
}

\author{Michael Davies}
\email{mdavies@nvidia.com}
\affiliation{%
  \institution{NVIDIA}
  \city{Santa Clara}
  \state{CA}
  \country{USA}
}

\author{Rahul Chatterjee}
\email{chatterjee@cs.wisc.edu}
\affiliation{%
  \institution{University of Wisconsin-Madison}
  \city{Madison}
  \state{WI}
  \country{USA}
}

\author{Karthikeyan Sankaralingam}
\email{karu@cs.wisc.edu}
\affiliation{%
  \institution{University of Wisconsin-Madison \& NVIDIA}
  \city{Madison}
  \state{WI}
  \country{USA}
}

\begin{abstract}

GPUs are the dominant platform for many important applications today including deep learning~, accelerated computing, and scientific simulation~. However, as the complexity of both applications and hardware increases, GPU chip manufacturers face a significant challenge: how to gather comprehensive performance characteristics and value profiles from GPUs deployed in real-world scenarios. Such data, encompassing the types of kernels executed and the time spent in each, is crucial for optimizing chip design and enhancing application performance. Unfortunately, despite the availability of low-level tools like NSYS and NCU, current methodologies fall short, offering data collection capabilities only on an individual user basis rather than a broader, more informative fleet-wide scale. This paper takes on the problem of realizing a system that allows planet-scale real-time GPU performance profiling of low-level hardware characteristics. The three fundamental problems we solve are: i) user experience of achieving this with no slowdown; ii) preserving user privacy, so that no 3rd party is aware of what applications any user runs; iii) efficacy in showing we are able to collect data and assign it applications even when run on 1000s of GPUs. Our results simulate a 100,000 size GPU deployment, running applications from the Torchbench suite, showing our system addresses all 3 problems.

\end{abstract}



\maketitle
\section{Introduction}
GPUs are the dominant platform for many important applications today including deep learning~\cite{mittal2019survey}, accelerated computing, and scientific simulation~\cite{9623445}. However, as the complexity of both applications and hardware increases (the recent Hopper GPU includes sophisticated components like a tensor memory accelerator, support for atomics, and the latest generation TensorCore~\cite{choquette2022nvidia,hopper-whitepaper}), GPU chip manufacturers face a significant challenge:  \emph{how to gather comprehensive performance characteristics from GPUs deployed in real-world scenarios}. Despite the availability of low-level tools like NSYS and NCU~\cite{ncu-counters}, current methodologies fall short, offering data collection capabilities only on an individual user basis rather than a broader, more informative planet-wide scale.

This paper argues for the necessity of a new paradigm in GPU performance data collection, envisioning a form of statistical debugging or continuous planet-wide profiling.  A key part of \textbf{our goal is to make this data available to chip developers as opposed to a hyperscalar's ``internal'' use and characterization.}


\emph{This paper envisions to collect low-level performance counter data from deployed GPUs across a large number of users while causing negligible interference to or detraction of the user experience. The recipients of this profiling data are \textbf{chip designers to improve} the performance of GPUs.}

Achieving this vision involves several key challenges, particularly in the areas of observability, privacy, and efficient data management. In this context, observability refers to which performance counters can be monitored and how accurately they reveal the running application. The data then should be collected by the chip designers for further analysis. This data collection and aggregation must be done in a way that preserve the privacy of the user applications. Existing solutions~\cite{opentelemetry,dynolog,codeguru,azure-monitor,datadog,pyroscope,parca,CLEMENTE2023126585,splunk} fall short in two major ways: they lack privacy-preserving methods for users to share performance data with \textbf{chip manufacturers}, and they do not enable manufacturers to effectively  analyze data from diverse sources without compromising user privacy.


\begin{figure}[t]
    \centering
    \includegraphics[width=\columnwidth]{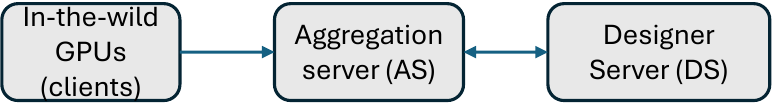}
    \caption{\sysname Components}
    \label{fig:intro-overview}
    \vspace{-0.2in}
\end{figure}

Addressing these aforementioned three challenges, this paper introduces a three-pronged solution designed to work seamlessly with state-of-the-art unmodified GPU hardware deployments and off-the-shelf software. Our overall system, called \sysname, is outlined in Figure~\ref{fig:intro-overview}. 
\sysname introduces an untrusted (honest-but-curious) aggregation server (AS) which receives encrypted data directly from GPUs, aggregates them, and only shares the encrypted aggregated data with the chip designers (designer servers or DS). The AS can be hosted on public cloud services such as Azure, GCP, AWS, or Equinix and run by an organization separate from DS. Second, our solution incorporates Additive Homomorphic Encryption (AHE)\footnote{Although fully homomorphic encryption allows performing any operation over encrypted data, it is computationally very expensive, while a restriction to addition provides a variety of other simpler and faster solutions~\cite{10.1145/3214303}.} --- through careful system design, only additions on encrypted data are needed to process data on untrusted third-party servers, ensuring that DS can gain insights without ever accessing raw user data directly, while preventing the AS from learning anything about the hardware performance counter data or individual users who are participating in \sysname.  We develop an automatic application identification and classification approach. \sysname uses extremely low sampling rates combined with wide-scale execution to achieve speed and coverage across applications.  To concretize the solution, we focus on workstation/data-center class GPUs, although the principles translate to other accelerators (e.g. accelerators on mobile devices' or autonomous vehicles' SoCs), FPGAs, and CPUs. 
In practice, we show that a single aggregation server on a m3.small.x86 Equinix instance with a 25~Gbps network link can easily service 100,000 GPUs --- the cost of such a deployment is approximately \$6000 per year~\cite{equinix}. Penrose's aggregated per-performance counter histograms are extremely useful to chip designers, they can examine how intensely different hardware units (e.g.,  tensor cores, caches, memory bandwidth) are exercised under the  applications run in-the-wild as opposed to benchmarks, or limited in-house workload characterization studies.

To evaluate \sysname, we used a hybrid approach that includes measurements on real hardware considering the entire Torchbench suite (150+ applications), an implementation of AHE aggregation using the Paillier library~\cite{intel-paillier}, and network traffic measurements. We built a \sysname planet-scale simulator that simulates GPU deployment at scale to study coverage, time to collect data, network effects, and cost effectiveness. This paper's contributions are:

\begin{figure*}[tbp]
    \centering
    \includegraphics[width=\textwidth]{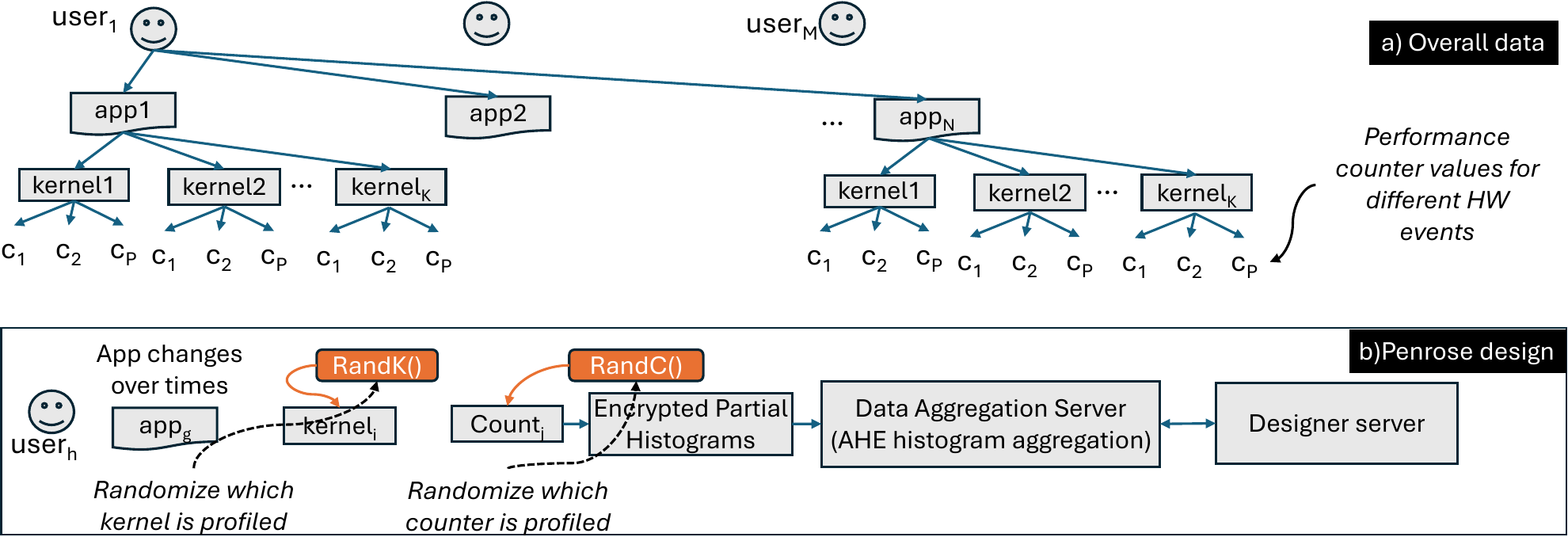}
    \caption{\sysname conceptual overview}
    \label{fig:concept-overview}
    \vspace{-0.1in}
\end{figure*}

\newcommand{\spacer}{\noindent$\bullet$\ \ }
\begin{itemize}
\item  Design of the \sysname system that allows at-speed planet-wide GPU performance profiling without any hardware modifications.
\item A novel histogram aggregation algorithm using partial homomorphic accumulation on encrypted data to provide user privacy and allowing only chip designers to access performance data.
\item A formalization of \sysname's privacy guarantees under a practical threat model we describe.
\item Simulation-based results that show \sysname is effective; it obtains over $99\%$ application coverage within 8~hours, when running across 100,000 GPUs, needing a single aggregation server, and has an average slowdowns of $0.045\%$. This is complemented by measurement on 32-GPU testbed with \~400 hours of GPU execution across 154 applications.
\item We demonstrate a chip designer use-case showing data collection across 8 SOTA LLMs. 
\end{itemize}

Paper organization: \S\ref{sec:overview} provides an overview of Penrose, \S\ref{sec:system} provides its detailed design, \S\ref{sec:methodology} discusses evaluation methodology, \S\ref{sec:results} presents our results, \S\ref{sec:related} discusses related work, and \S\ref{sec:conc} concludes.

\mypagebreak

\section{\sysname Overview}\label{sec:overview}
We present \sysname, a system designed to profile GPUs at a global scale. An ideal system would capture performance counter distributions for every kernel of every application running on end-user systems, without perceptible slowdowns or privacy breaches. This data, aggregated into histograms for analysis by a chip designer server (DS), is sourced from consenting GPU users and aggregated by a trusted aggregation server (AS). Figure~\ref{fig:concept-overview}(a) provides a conceptual overview of this data flow.

\subsection{Why Performance Telemetry} Such a planet-scale system is necessary because existing data sources are insufficient for guiding future hardware design for chip designers. Looking at DL, while valuable, open-source models do not reflect the behavior of widely deployed, closed-source systems like GPT-5 or Gemini. Looking ahead, the diversity of GPU applications is projected to grow substantially, encompassing a vast ecosystem of workloads beyond open-source deep learning. Similarly, hyperscaler profiling tools like DCGM are siloed, lack privacy guarantees, and are not generalizable across organizations, \textbf{and don't provide data to chip designers}. \sysname addresses this gap by creating a viable path for collaborative, privacy-preserving telemetry. By capturing the true diversity of real-world workloads, it enables a data-driven feedback loop between end-users and chipmakers, grounded in a well-established practice of opt-in, privacy-focused data collection seen in modern operating systems and services. The principle of opt-in telemetry is already a well-grounded practice in modern computing, from operating systems~\cite{leith2021mobile}, chips diagnostics(Intel PMT~\cite{intel-pmt,10255554}, GeForce Experience~\cite{geforce-experience}),  to vehicle hardware~\cite{telsa-telemetry}. Penrose builds upon this established model by introducing strong cryptographic privacy, ensuring the collected data can guide future hardware design without compromising user anonymity. \S \ref{subsec:usingpenrose} includes a description of how chip designers could use Penrose's telemetry.

\subsection{Strawman Solution and Challenges}\label{sec:strawman}
A strawman approach using standard tools like NVIDIA's NSight suite to collect and transmit performance data from all users would immediately fail due to four fundamental challenges. \textbf{First, user slowdown:} Comprehensive profiling incurs prohibitive $2\times$ to $10\times$ application slowdowns. \textbf{Second, data overload:} Continuous data capture from millions of users would create unmanageable network and storage burdens. \textbf{Third, privacy infringement:} Transmitting plaintext kernel and application names would violate user privacy. \textbf{Fourth, application identification:} The GPU driver only sees a stream of kernels, making it difficult to associate performance data with the correct high-level application.

\subsection{GPU Application Identification}
To overcome the lack of a reliable application identifier, we introduce the concept of a "snippet": a sequence of dynamic kernel name invocations of a fixed length. To disambiguate kernels that may share the same string name, we define a kernel’s identity not by its string name but by a cryptographic hash of its binary content (CUBIN/PTX). The GPU driver has access to this binary, making this hash trivial to compute. Throughout this paper, when we refer to a ``kernel name,'' we are referring to this unique, content-based hash. A snippet acts as a fingerprint for an application. In practice, to preserve privacy and enable efficient comparison, this kernel sequence is converted into a locality-sensitive Min-Hash (a \textbf{Snippet Sequence Min-Hash}) for similarity matching. A final cryptographic hash of this Min-Hash (a \textbf{Snippet Hash}) is then used for exact and anonymous lookups at the designer server. This allows \sysname to group performance data by application without ever knowing the application's actual name.

\subsection{Privacy Requirements}\label{sec:privacy-req}
Our threat model assumes honest-but-curious parties, where the AS and DS follow the protocol but may attempt to glean extra information; they are assumed not to collude. Our design provides three core guarantees. \textbf{(1) User Anonymity:} The AS and DS cannot determine if a specific user contributed to an aggregated histogram. \textbf{(2) Application Confidentiality:} The kernel sequence of a user's private application is never revealed to the AS or DS. \textbf{(3) Histogram Confidentiality:} The AS only handles encrypted histograms, and the DS only ever learns final, aggregated histograms from a population of users. The AS and DS are only permitted to learn the frequency of an anonymized application, identified by its Snippet Hash.

\subsection{The \sysname{} Solution}
\sysname solves the four core challenges with a scalable, privacy-preserving framework, shown in Figure~\ref{fig:concept-overview}(b). To prevent \textbf{user slowdown} and handle \textbf{data overload}, it uses randomized sparse sampling on each client, relying on the law of large numbers across many users to build complete profiles over time. Data is aggregated locally into histograms before transmission, dramatically reducing data volume. To ensure \textbf{privacy}, kernel names are hashed, and all partial histograms are encrypted with additive homomorphic encryption (AHE) before being sent to the AS over an anonymity network like Tor. Finally, for \textbf{application identification}, it uses the snippet-based fingerprinting described previously to group performance data by application.

To minimize performance overhead while avoiding sampling bias, \sysname uses a randomized approach. A kernel is profiled every \textbf{sampling interval} ($S$), and this interval is reset every \textbf{sampling reset interval} ($O$) seconds. On the client, once an \textbf{aggregation threshold} ($A$) of samples is collected for an application, they are compiled into \textbf{partial histograms}. These are then encrypted using an additively homomorphic encryption (AHE) scheme and transmitted to the aggregation server (AS). The AS combines these into an \textbf{aggregated snippet histogram (ASH)} for each application snippet. Finally, the AS sends the ASH data to the designer server every \textbf{server report interval} ($\delta$) minutes. Table~\ref{tab:design_vars} (in page~\pageref{tab:design_vars}) describes all the parameters used by \sysname.

The efficacy of \sysname relies on two key assumptions: a large number of participating users and the existence of ``hot'' (popular) applications, which are necessary to achieve good coverage with low-rate sampling. \sysname is designed to support distributed workloads by profiling each GPU independently. In settings like tensor parallelism, it effectively captures local kernel sequences, and its privacy-preserving aggregation remains scalable. While correlating snippets across GPUs is a future engineering goal, it is orthogonal to Penrose's primary contribution: the secure aggregation of local telemetry, a capability already validated under conditions more stringent than those of typical distributed workloads. Our prototype is built for NVIDIA hardware, but its core principles are readily adaptable to other platforms. 

\newcommand{\snippet}{\mathbf{K}}

\begin{figure}
    \centering
    \includegraphics[width=\columnwidth]{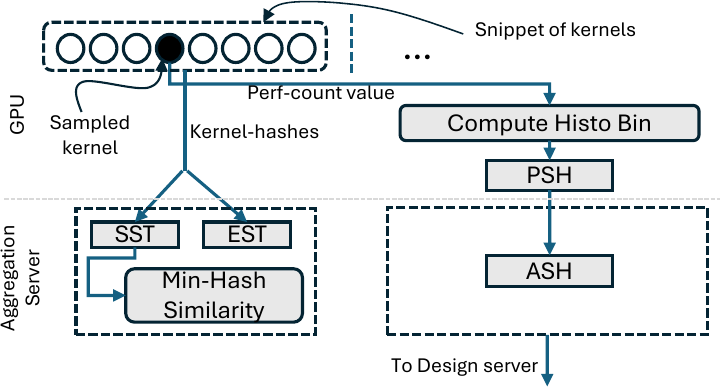}
    \caption{Dataflow from users to aggregation server.}
    \label{fig:dataflow}
    \vspace{-0.2in}
\end{figure}

\section{\sysname System Architecture}\label{sec:system}
The \sysname system operates in two logical phases: snippet classification (\S\ref{subsec:snippetclassification}) to identify applications, and data collection (\S\ref{subsec:datacollection}) to gather performance metrics. Figure~\ref{fig:dataflow} illustrates the information flow between system components. Appendix~\ref{app:protocol} provides the formal specification of the protocol and a proof sketch.

\subsection{Snippet Classification}\label{subsec:snippetclassification}
Application identification begins on the client, which observes a sequence of kernel invocations. Once a sequence reaches a predefined length $L$, it forms a snippet $\snippet = (k_1, k_2, \dots, k_n)$. To preserve privacy, the client does not transmit the raw kernel names. Instead, it computes a MinHash signature over the snippet's overlapping 8-grams. For a set of 8-grams $G(K)$ from a kernel sequence $K$, the MinHash is:
\[
\func{MinHash}(K) = \left( \min \{h_j(g) \mid g \in G(K)\} \mid j = 1, 2, \dots, H \right)
\]
where we use $H=100$ hash functions. The client sends this \textbf{Snippet Sequence Min-Hash} and its corresponding SHA-256 hash (the \textbf{Snippet Hash}) to the Aggregation Server (AS).

The AS uses the Snippet Hash for a fast lookup in its Equivalent Snippet Table (EST) to map the incoming snippet to a known \textbf{Canonical Snippet}. If no exact match is found, the AS uses the MinHash signature to find the closest match in its Snippet Sequence Table (SST) via Jaccard similarity. The Jaccard similarity $J(M_i, M_j)$ between two MinHash signatures is estimated as:
\[J(M_i, M_j) = \frac{|M_i \cap M_j|}{|M_i \cup M_j|}.\]
If $J$ exceeds a threshold $\tau=0.85$, the snippets are considered to be from the same application.

\subsection{Data Collection Phase}\label{subsec:datacollection}
\paragraph{Client-Side Sampling and Aggregation} To minimize overhead, \sysname samples every $N$th kernel instance, with the starting offset periodically randomized to ensure statistical coverage. We only collect metrics that do not require kernel replays (Appendix~\ref{app:counters} has partial list for A100 GPUs) and can collect pairs of counters to study correlations by repurposing a 128-bin histogram as a 32x32 2D histogram. Sampled performance data is aggregated locally on the client into a 128-bin \textbf{Partial Snippet Histogram (PSH)}. Once an aggregation threshold is met, the PSH is encrypted using the Paillier additive homomorphic encryption (AHE) scheme with a 2048-bit modulus and sent to the AS.

\paragraph{Statistical Justification of Kernel Sampling Coverage}
Let an application execute $k$ kernels per run. Each user samples one kernel every $N$ executions, with the starting offset drawn uniformly at random from $[0, N-1]$. Over $u$ independent users or runs, the probability that a specific kernel instance at index $i$ is not sampled in a single run is:$P_{\text{miss}} = 1 - \frac{1}{N}$. The probability that it is \emph{never} sampled over $u$ runs is:$P_{\text{never}} = \left(1 - \frac{1}{N}\right)^u$. Thus, the probability that kernel $i$ is sampled at least once across $u$ independent users is:$
P_{\text{hit}} = 1 - \left(1 - \frac{1}{N}\right)^u$. For example, with $N = 100$ and $u = 1000$ users, we have: $P_{\text{hit}} \approx 1 - e^{-10} \approx 1 - 4.5 \times 10^{-5}$, showing that nearly all kernel instances will be covered with high probability. This analysis holds for any deterministic or nondeterministic kernel sequence and generalizes to workloads with variable-length runs. In Section~5, we empirically examine this behavior.

\paragraph{Server-Side Operations}
The AS uses the homomorphic property of Paillier encryption to add newly received encrypted PSHs to the corresponding \textbf{Aggregated Snippet Histogram (ASH)} for that application snippet, without ever decrypting the data. Periodically, the Designer Server (DS) retrieves these encrypted ASHs from the AS and decrypts them using its private key for analysis.

\paragraph{Extensions} \sysname{} has been designed with chip designers receiving the telemetry. It can be deployed and used by others like hyperscalars or data-center operators. Another extension is to use Penrose to aggregate differently: per-kernel histograms aggregated across all users. 
The DS can further log data internally as part of their telemetry pipeline to perform analytics on time varying behavior etc. Finally, the DS might desire data from different chip SKUs aggregated differently, whose trivial implementation is to dedicate an AS per SKU, or have the AS listen on different ports for each chip SKU. 

\subsection{Protecting User Privacy}\label{sec:privacy}
\sysname's privacy model is built on a combination of cryptographic techniques chosen to balance utility, efficiency, and robust protection under an honest-but-curious threat model. Our core design combines Additive Homomorphic Encryption (AHE) with an anonymity network. We explored several alternatives before finalizing this approach. Fully Homomorphic Encryption (FHE), which supports arbitrary computation, is unnecessarily powerful and inefficient for our needs, as the AS only performs summation. We also carefully considered various forms of Differential Privacy (DP), such as LDP~\cite{ldp} and shuffle-DP~\cite{shuffleDP}. While DP offers strong, formalized privacy budgets against reconstruction and membership inference attacks, it requires injecting noise into the performance data. Our experiments showed this noise unacceptably degrades the utility of the fine-grained counter distributions we aim to measure.

Instead, our system uses Tor for anonymity, which acts as a practical shuffle step, and end-to-end AHE to protect data confidentiality. Furthermore, our collected data consists exclusively of integer streams without the auxiliary metadata or relational tables that are often exploited in correlation attacks~\cite{shmatikov,grubbs}. Thus, while we recognize the value of DP in certain contexts, it is not directly applicable to or necessary for our specific design.

\paragraph{1. Histogram confidentiality} \sysname ensures that AS cannot learn anything about the histograms performance counter (that they do not already know, like the number of buckets in each histogram, the number of performance counters in the telemetry data, etc.). To achieve this, we use an additive homomorphic encryption (AHE)~\cite{paillier} to encrypt the partial histograms so that AS cannot learn anything about the partial histograms but is able to compute the aggregation (summation over each histogram bucket). We use the Paillier's AHE scheme~\cite{paillier}.

DS selects a public-private Pailler key pairs during the setup of \sysname; DS publicly post the public keys (with certificates) on their website, from where users can download when they opt into \sysname.   All users use the same public key. Paillier encryption of a message $c_i$ --- a histogram value in our case --- with a public key $pk=(g, n)$ is defined as follows:  $
   \text{Enc}_{pk}(c_i) = g^{c_i} \cdot r^{n} \mod n^2$,
   where 
   \( r \) is a random number selected per encryption. After receiving encrypted histograms, the AS can aggregate them without decryption using Paillier’s homomorphic property:    $\text{Enc}_{pk}(c_1) \cdot \text{Enc}_{pk}(c_2) = (g^{c_1} \cdot r_1^{n} \mod n^2)\cdot(g^{c_2} \cdot r_2^{n} \mod n^2) = (g^{c_1+c_2} \cdot (r_1\cdot r_2)^{n} \mod n^2) = \text{Enc}_{pk}(c_1 +c_2)$. Thus, AS can simply multiply the encrypted values in each histogram bucket from different updates for the same snippet hash for the same performance counter to create the ciphertext of the sum of the underlying histogram values.  Only DS who has the private key can decryt these ciphertexts. When DS receives the encrypted aggregated histograms, they  decrypt the aggregate ciphertexts to obtain the aggregate snippet histogram, but cannot learn about the partial histograms. 

In our implementation, we use a public key size of $2048$-bits, which can encode  any $64$-bits value and operate on it without the worry of any overflow even after trillions of additions. NIST recommend $2048$-bit keys (Table~2 in~\cite{nist}) as  provide sufficient security (equivalent of $112$-bits) for most applications. 

\paragraph{2. User Anonymity}
An attacker's goal is to link a specific user (e.g., via IP address) to their submitted data. To prevent this, clients route their updates through an anonymity network. While several options exist, such as Mixnet~\cite{chaum1981untraceable}, Loopix~\cite{203838}, or Apple's iCloud Private Relay~\cite{sattler2022icloud,apple2023icloud}, we use Tor~\cite{dingledine2004tor} in our implementation due to its widespread availability and vetting. Crucially, a new Tor circuit is established for each transmitted histogram, preventing the AS from linking multiple submissions over time. The combination of an anonymity network and AHE (which blinds the histogram content) effectively severs the link between a user and their data. Tor might be vulnerable to Sybil attacks~\cite{levine2006survey}, that can violate the anonymity of end users. However, solving Sybil attack requires different set of techniques that is orthogonal to the goal of this paper, and Tor already has a number of mechanisms to reduce the risk of such attacks~\cite{levine2006survey}. In \secref{sec:results}, we demonstrate that the bandwidth requirement for \sysname is small (2MB/sec across 100,000 GPU nodes) and can be handled easily by existing Tor networks. 

\paragraph{3. Application confidentiality} We want to ensure that AS or DS cannot learn the sequence of kernels of a private application a user is running. We consider the applications private for which AS or DS do not have access to the code or binary executable.  For public applications, AS can learn the snippet sequence from the code or application binaries even without any update from the user. AS can use tools like {\tt nsys} and other NSIGHT tools~\cite{nvidia_nsight_systems} to obtain a trace of kernels and use our public \sysname code to compute the snippet min-hash. Thus, we do not aim to protect the kernel sequence of public applications.  

Thus, \sysname never sends the raw snippet --- sequence of kernel names --- to AS.
Instead, \sysname sends snippet hash and snippet min-hash values computed using a secure cryptographic hash function such as SHA-256. 
From the snippet min-hash, AS learns the smallest hash value among all $n$-grams of kernel sequence for each hash function $h_1, h_2, \ldots, h_H$. As \sysname uses cryptographic hash functions, the only way to uncover the underlying $n$-gram from the hash values via \emph{guess-and-check} approach used in password cracking~\cite{hranicky2019distributed,lubeck2013johntheripper}. 
We show next that using such techniques are infeasible in our context due to modest alphabet size (number of unique GPU kernel names) and $n$-gram sizes we use, $n=8$. 

There are approximately $N=10^4$ publicly available NVIDIA GPU kernels~\cite{cudakernels}. Thus there are $N^n$ unique kernel $n$-grams are possible. 
To uncover a hash value, AS has to bruteforce search in this large search space $N^n$ kernel $n$-grams.  We select $n=8$ to make this computation significantly harder; and for $N=10^4$, we have $N^8 \approx 10^{32}$. The complete Bitcoin networks can compute about $10^{21}$ hashes per second~\cite{cointribune2024bitcoin}. That means to break the hash value of an $8$-gram, an adversary even with the compute power equivalent of the complete Bitcoin network in the world will take over 3,100 years.  Although it is computationally hard to identify \textit{any} kernel 8-gram, an attacker could identify a set of popular 8-grams and learn their presence in an application. 

To avoid such kernel $n$-gram identification, we propose to include a \textbf{per-application salt} selected by the \textit{developer} during the application (or library) compilation for all users of that application (or library).\footnote{At present CUDA compiler, as far as we know, does not allow the CUDA kernel name to be stripped or obfuscated.} This can be done by GPU companies providing this as a compiler option (with developers concerned about application anonymity deploying their applications with this flag used during compilation). Or it can be done with a trivial source-to-source compiler that inserts a wrapper function name with an obfuscated name. For cases where the GPU chip company ships libraries in pre-compiled form, the developer can use a binary rewriter that mangles kernel names of these libraries before deployment. 
The telemetry capability we add needs to retain this kernel name anonymity. Our name mangling solution described above achieves this.

\paragraph{Additional privacy considerations/Extensions} In our threat model we considered all parties to be honest but curious. Malicious users could submit fake partial histogram data to AS to subvert the accuracy of the analytics system (i.e. these poisoning attacks are well known in ML~\cite{poisoning} and techniques to combat these attacks can be readily layered on top of Penrose). One approach for handling poisoning is that the client proves using ZKPs that the data under encryption satisfies a validity predicate (called input validation), such as Eiffel and ACORN~\cite{Eiffel,ACORN}. However, these solutions incur too much overhead in our context. Hardware attestation like the one available in the latest generation of data-center class GPUs like Hopper could also be used to verify the veracity of the partial histograms. H-100 GPUs provide device identity certificate signed with a device-unique ECC-384 key pair~\cite{h100-attestation} and can sign the performance counters that can be verified that the data is indeed generated on a legitimate GPU. Other similar techniques such as anonymous attestation~\cite{brickell2004direct, brickell2008new,camenisch2016anonymous} or enhanced privacy ID (EIPD) introduced by Intel~\cite{johnson2016intel,brickell2007enhanced} could be used as well. Moreover several prior works provide approaches to transform semi-honest protocols to a protocol that is secure in the fully malicious model (see~\cite{Katz-Lindell} and \cite{boneh-shoup}) can be applied to \sysname as a mitigation, with some optimizations needed to reduced their overheads.
\section{Evaluation Methodology}\label{sec:methodology}
\begin{figure}
    \centering
    \includegraphics[width=\columnwidth]{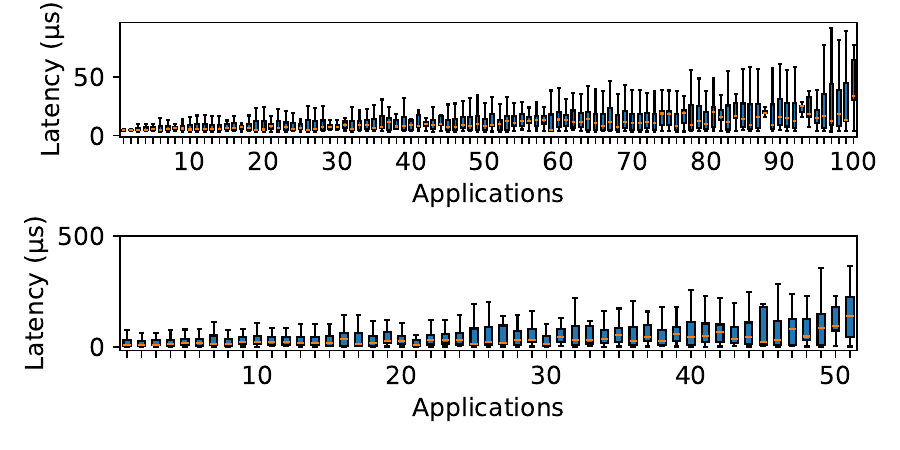}
    \caption{Latencies of kernels per application for 154 applications we tested. 3 applications whose max kernel latencies exceeded 500$\mu s$ are not shown here. X-axis is application number, with the ``small'' 100 shown in the 1st plot.}
    \label{fig:kernel-lat}
    \vspace{-0.2in}
\end{figure}

We evaluate \sysname using a custom, event-driven system simulator designed to model its planet-scale behavior. The simulator models a configurable number of GPUs, the aggregation server (AS), and the designer server (DS), including the network traffic between them and the computational overhead of cryptographic operations. It faithfully implements our sampling strategy, including randomized offsets and user activity load factors, allowing us to project long-run system performance, such as the time required to achieve full application coverage.

The simulator is driven by real-world kernel traces from 154 Torchbench applications~\cite{hao2023torchbench,torchbench2}, captured from an NVIDIA A100 GPU using Nsight Systems (NSYS). These applications exhibit a wide diversity in kernel latencies (3$\mus$ to 521$\mus$) and complexity (14 to 128,838 kernels per batch), providing a realistic and challenging workload.

To ensure our simulator's performance model is grounded in reality, we augment it with direct measurements from a prototype of \sysname's core components. We measured the latency of AHE histogram aggregation using Intel's Paillier library and the performance of our MinHash-based application identification. We also measured the true user slowdown of sampling on an A100 GPU with Nsight Compute (NCU) at various rates.

Finally, we validated our simulation framework against two physical deployments. First, we deployed \sysname on a 32-node A100 GPU testbed to confirm its behavior on real hardware. Second, we used CloudLab to validate the network scalability, running a full-stack prototype with 100 physical clients emulating the traffic of 100,000 concurrent GPUs. This deployment, capturing all communication and cryptographic overheads, confirmed that our simulator accurately projects the system's runtime behavior at scale.
\section{Evaluation}\label{sec:results}

\begin{table}[t]
\small
\centering
\begin{tabular}{llrl}
\toprule
Var. & Description & Default & Units \\
\midrule
$L$ & Snippet length & 10,000 & \texttt{Kerns} \\
$S$ & Sampling interval & 10,000 & \texttt{Kerns} \\
$O$ & Sampling reset interval & 600 & \texttt{Seconds} \\
$A$ & Aggregation threshold & 10,000 & \texttt{Samples} \\
$\delta$ & Server report interval & 1 & \texttt{day} \\
$G$ & Users per server & 100,000 & \texttt{GPUs / Server} \\
\bottomrule
\vspace{2pt}
\end{tabular}
\caption{Different parameters used by \sysname.}
\label{tab:design_vars}
\vspace{-0.35in}
\end{table}

We evaluate \sysname's effectiveness and deployment costs we use the default parameters in Table~\ref{tab:design_vars}. To clarify scope, all results on profiling accuracy, overhead, and slowdown (Figures~\ref{fig:kernel-lat}, \ref{fig:app-slowdown},\ref{fig:small-gpu-runs},\ref{fig:histo-error}, Tables~\ref{tab:snippetaccuracy}, \ref{tab:crypto-speed}) are from direct measurements on NVIDIA A100 GPUs. Only the planet-scale convergence-time experiments (\S\ref{subsec:coverage}) use our validated discrete event simulator.

\subsection{Measuring User Slowdown}

\begin{figure}
    \centering
    \includegraphics[width=\columnwidth]{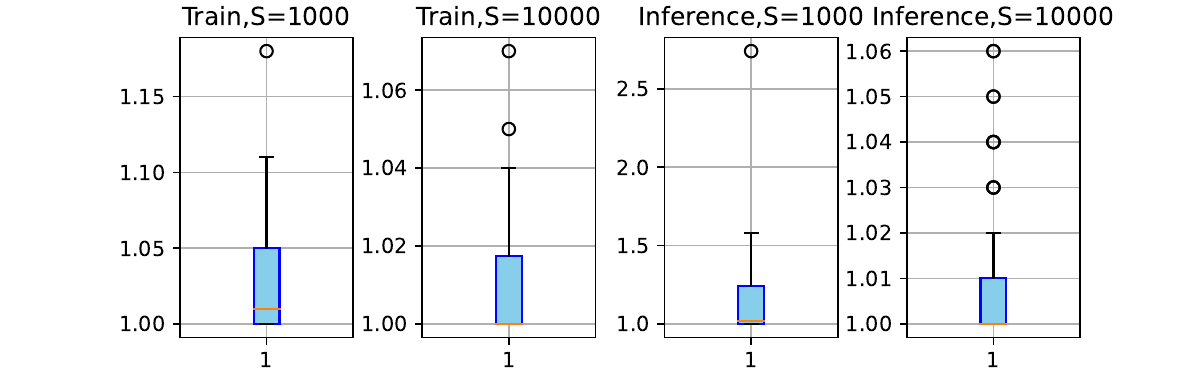}
    \caption{Application slowdown from ncu for Torchbench for 1/1000 and 1/10,000 sampling for Train and Inference.}
    \label{fig:app-slowdown}
    \vspace{-0.2in}
\end{figure}

Reading GPU performance counters introduces overhead, which we mitigate by sampling every $S$-th kernel. Figure~\ref{fig:app-slowdown} shows the application-level slowdown for sampling intervals ($S$) of 1,000 and 10,000. While $S=1000$ incurs a noticeable 7.55\% average slowdown, increasing the interval to $S=10,000$ reduces the average slowdown to just 0.045\%, with a maximum of 7\% for short-running applications. We therefore select $S=10,000$ as our default.

\emph{\textbf{Takeaway:} With a sampling interval of 10,000, \sysname imposes a negligible average slowdown of 0.045\% across 154 real-world DL applications.}

\subsection{Application Coverage}\label{subsec:coverage}
\subsubsection{Large scale Simulation} 
\begin{figure*}
    \centering
    \includegraphics[width=0.325\textwidth]{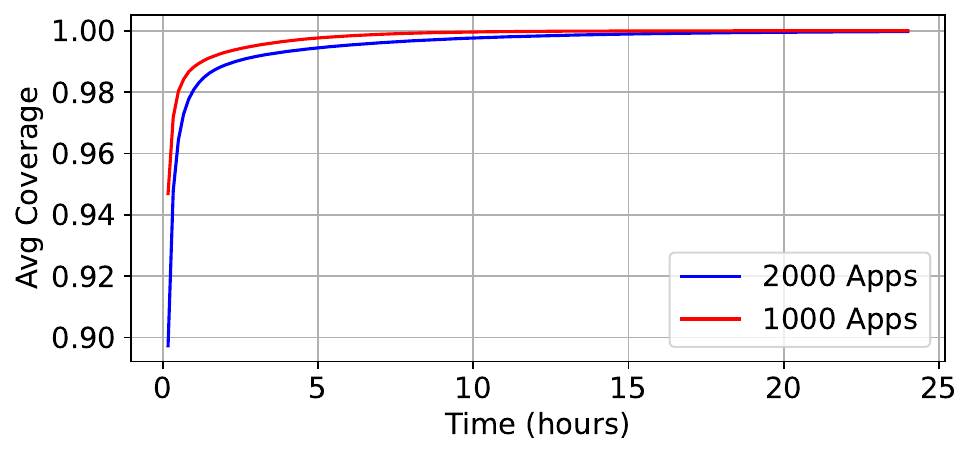}
    \includegraphics[width=0.325\textwidth]{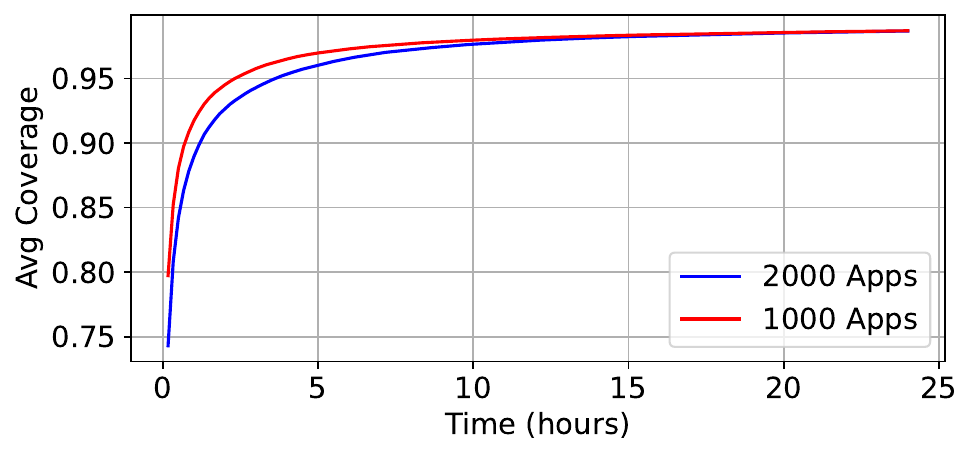}
    \includegraphics[width=0.325\textwidth]{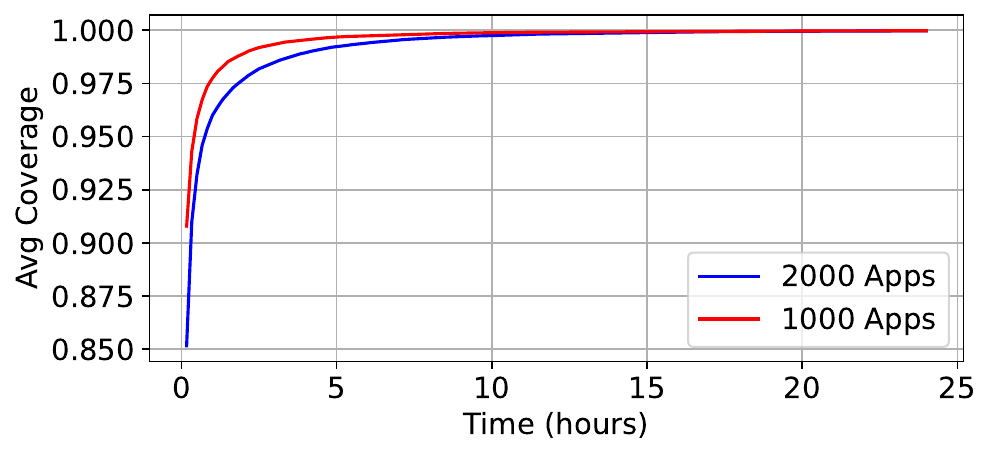}\\

    \includegraphics[width=0.325\textwidth]{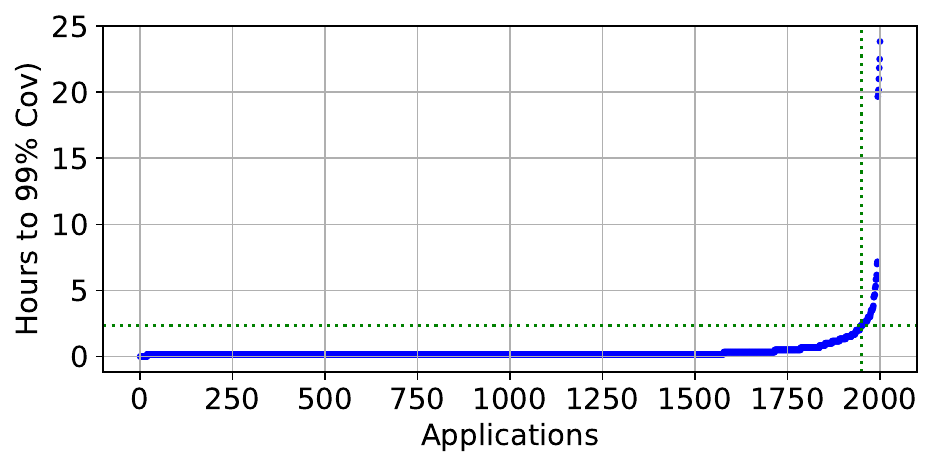}
    \includegraphics[width=0.325\textwidth]{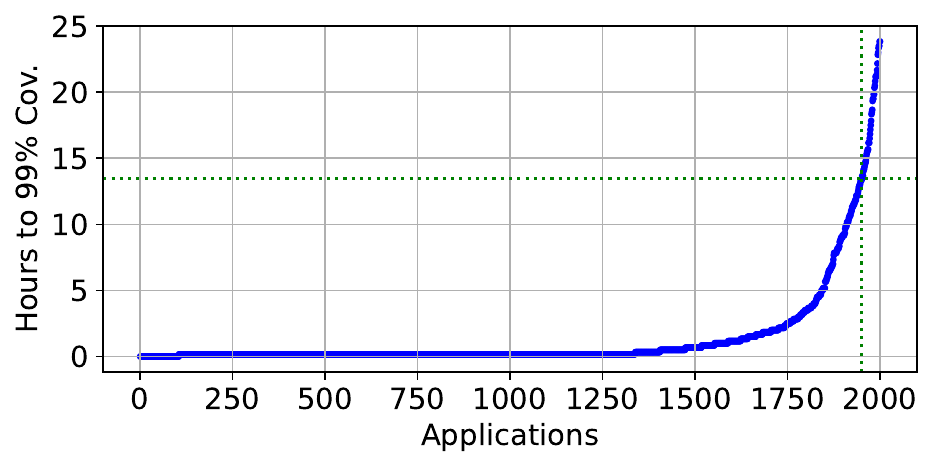}
    \includegraphics[width=0.325\textwidth]{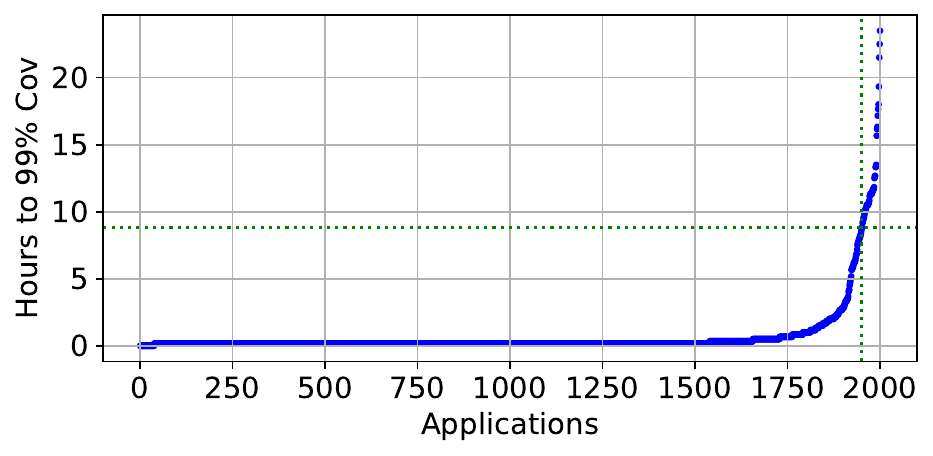}\\
    \small{(\textbf{left}) Uniform distribution (\textbf{middle}) Normal Dist (small apps most frequent) (\textbf{right}) Normal Dist (large apps most frequent)}
    \caption{Average coverage (1000 and 2000 apps) and Time to 99\% coverage (2000 apps) averaged across all simulated applications for 100,000 GPUs.  The green vertical line marks the 97.5\%-th percentile (1950 applications). Load factor is 10\% for all scenarios.}
    \label{fig:cov_line}
\end{figure*}

Sampling kernels for performance monitoring introduces a new problem: how to ensure that performance data of all kernels are considered in the aggregated histogram to accurately represent application-level performance. We define ``full'' coverage for an application as having recorded data for a performance counter for all of its kernels. Coverage for one application, then, is expressed as a fraction of kernels whose data has been reported to the \CS{}. If we run 154 applications on 100,000 GPUs, we get full coverage in a few minutes. Across Torchbench, average kernel execution time is only 30$\mu s$, and with an uniform distribution every application has 600 copies running simultaneously. For this experiment, we instead simulate 2000 unique applications built from this list of 154 --- each application has 13 copies, each of which is treated as a unique application.
We also take this application set of 2000 and create two normal distributions, In {\tt normal-large}, the applications with the longest number of kernels are most frequently executed by users. In {\tt normal-small}, the applications with the fewest number of kernels are most frequently executed by users. In both cases, the distribution has a mean of 1000 and std-deviation of 333. This implies that with {\tt normal-large}, 11.9\% of the applications are the top-200 largest, 38\% of the applications are the largest 660 applications, and 68\% are the largest 1320 applying standard probability calculations.   Real world application mix will fall in-between these two normal distributions. Figure~\ref{fig:cov_line} shows the \textit{average} application coverage of both $1000$ and $2000$ applications as a function of time (hours) to demonstrate how coverage improves as the ratio of users to applications increases in all 3 cases. We can see that for our canonical scenario, we easily achieve $>99\%$ coverage in 8 to 24 hours.  If a designer was interested in 10 performance counters, they could chose to wait 10 days for $>99\%$ coverage, or get 85\% of the applications $\times$ 10 counters covered every day by controlling when they switch which performance counters are sampled. We also report the amount of time each application took to reach $>99\%$ coverage in the scatter plot. The green lines show the time taken for 97.5\% of the applications (1950 of 2000) to reach $>99\%$ coverage. 

We studied and simulated systems of different scale (10,000 GPUs to 1 million GPUs), while keeping $S$ constant. With a million GPUs, \sysname achieves effectively the same coverage trends with 20,000 applications. The  curves are indistinguishable on the graph.

\emph{\textbf{Takeaway:} 
Based on our simulation, we find that high coverage is easily achieved when the number of participants ($G$) is large even with seemingly low sampling rate of $S=10,000$. Across a deployment of 100,000 GPUs, we find the system can still achieve $>99\%$ coverage in a day even when there are up to 2000 popular applications running in the wild, scaling to 20,000 applications with 1 million GPUs.}

\begin{table}[tbp]
    \centering
    
    \begin{tabular}{lcccccc}
    \toprule
            \multirow{2}{*}{\# APPS} & \multicolumn{3}{c}{\textbf{100,000 GPUs}} & \multicolumn{3}{c}{\textbf{10,000 GPUs}} \\
        \cmidrule(lr){2-4} \cmidrule(lr){5-7}
        & {U} & {$N_s$} & {$N_l$} & {$U$} & {$N_s$} & {$N_l$} \\
        \midrule

2000 & 2.3 & 13.5 & 9.5 & 15.3 & 14.5 & 12.7 \\
        1500 & 2.3 & 11.8 & 7.5 & 13.7 & 16.7 & 9.2 \\
        1000 & 1.5 & 10.5 & 4.8 & 10.2 & 13.2 & 8.8 \\
        500 & 0.7 & 5.2 & 2.3 & 6.7 & 15.2 & 10.2 \\
        200 & 0.2 & 2.3 & 1.0 & 2.2 & 11.3 & 11.7 \\

        \bottomrule
    \end{tabular}
    \caption{Hours to cover 97.5\% of apps.}\label{tab:varying-gpus}
    \vspace{-0.275in}
\end{table}

\subsubsection{Sensitivity to system scale} Table~\ref{tab:varying-gpus} examines time-to-coverage under different deployment scales and application distributions. We vary the number of GPUs (from 100,000 to 10,000) and the number of distinct applications (from 2000 to 200), across three distribution models: \textsc{Uniform ($U$} (apps appear with equal probability), \textsc{$N_s$} (apps with few kernels dominate), and \textsc{$N_l$} (apps with many kernels dominate). We report the time (in hours) required to achieve 99\% snippet coverage for 97.5\% of applications. As expected, fewer GPUs increase time to convergence, but the impact is modest: even with a 10$\times$ reduction in GPUs, time increases by only $\sim$2--3$\times$. Distribution shape also influences convergence—skewed workloads dominated by small kernels ($N_s$) take significantly longer to cover; since the apps with many kernels appear infrequently. As we decrease the number of applications, convergence happens faster. These results confirm Penrose remains effective across varying deployment sizes and workload skews.

\subsubsection{Testbed measurement} To address concerns about relying on simulation and to validate our approach on physical hardware, we deployed Penrose on a 32-node A100 GPU testbed within our internal cluster. In this deployment, we developed a wrapper script to select \textit{N} applications from the Torchbench suite. Each GPU repeatedly executed its assigned application. The applications were profiled using `ncu` with a sampling rate of 1/10000. We conducted 7 cohorts of experiments, varying the number of concurrent applications (*N*) from 1 to 32 in powers of two. For each cohort, we ran up to 30 trials, with each trial selecting a random set of \textit{N} applications; for the 1 application setting we ran 154 traisl. To ensure a diverse mix, one-third of the trials for each cohort used a normal-small distribution for application selection, one-third used a normal-large distribution, and the final third used a uniform distribution. In total, we executed 382 distinct experiments over approximately 20 days, accumulating 470 hours of total GPU execution time. The longest experiment (32 apps) ran for 21 hours.

\begin{figure}
    \centering
    \includegraphics[width=1.0\linewidth]{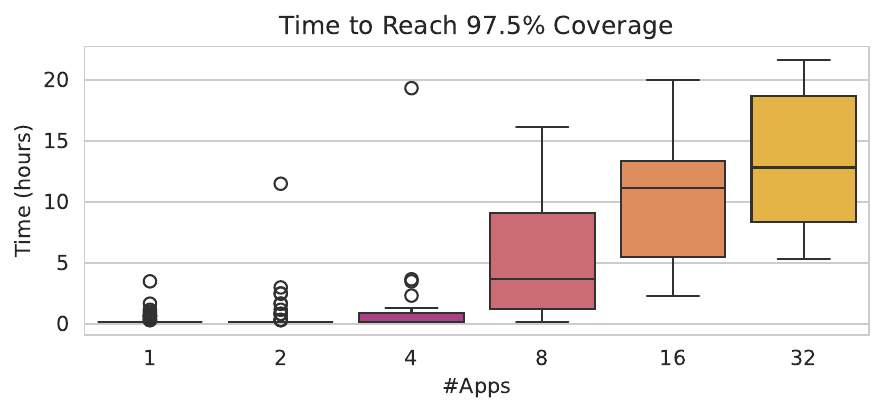}
    \caption{Time for convergence on GPUs. Small experiment.}
    \label{fig:small-gpu-runs}
\end{figure}

Figure~\ref{fig:small-gpu-runs} presents a box plot illustrating the time to achieve convergence 97.5\% for these experiments as the number of applications is varied. As one might expect, at a low number of applications, there is considerable variance in convergence time, depending on whether a small or large application was chosen. At a scale of 8 to 64 applications, the system consistently achieved full coverage within 4 to 12 hours. Most importantly, this large-scale experiment demonstrates Penrose's capability to collect data effectively on real hardware. To validate our simulator, each of these hardware application mixes was executed within our discrete-event simulator. The simulator produced convergence times identical to the hardware measurements. While a perfect match might seem surprising from the perspective of traditional cycle-level simulation, it is the expected outcome for our system's level of abstraction. This deterministic result occurs because both the physical deployment and the simulation are driven by the same sequence of discrete events: intermittent messages sent from each GPU to the aggregation server. With a sampling interval of 10,000 seconds and aggregation of 10,000 samples, the worst case is a message every 530 seconds for the application {\tt soft\_critic} which has avg kernel execution time of 5.3$\mu s$. Any real-world timing ``jitter'' in message arrivals—on the order of milliseconds or seconds—is orders of magnitude smaller than our final measurement granularity, which is in 100s of seconds. The encryption time is measured on the order of 100ms to 400ms. The simulation, therefore, correctly models the exact sequence of events that dictates convergence, leading to an identical result. It is important to note that as the system scales to thousands of GPUs, the primary challenge becomes managing network effects on the server, which we discuss later in this section. The load on individual clients, however, remains constant regardless of system size.

A crucial aspect of our methodology is quantifying the measurement error introduced by sampling. To this end, we conducted a rigorous validation experiment. First, we established a ground truth by generating "perfect" histograms of DRAM memory bandwidth for 154 applications, running each on a single GPU to capture its complete execution profile. We then generated corresponding sampled profiles by running each application across 32 GPUs, collecting data at a 1/10000 sampling rate, and aggregating the results. By comparing the sampled histogram to the ground truth for each of the 128 bins per application, we computed the relative error across a total of 19,712 data points.

\begin{figure}
    \centering
    \includegraphics[width=1.0\linewidth]{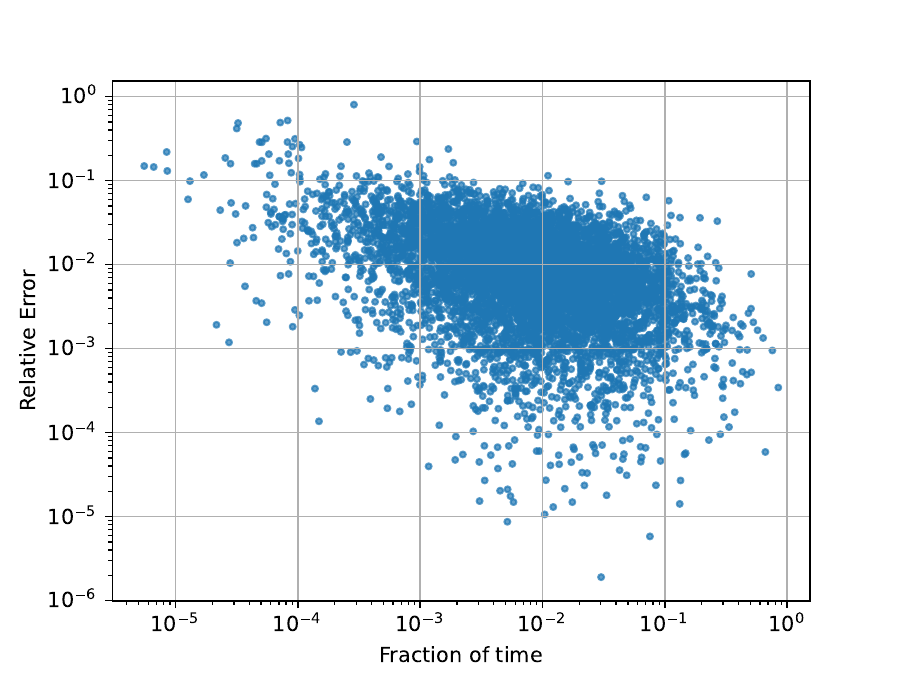}
    \caption{Relative errors in histograms.}
    \label{fig:histo-error}
\end{figure}

Figure~\ref{fig:histo-error} visualizes the results as a scatter plot, with the fractional execution time represented by a bin on the x-axis and the corresponding relative error on the y-axis (note both axes are on a log scale). \textbf{The plot clearly demonstrates that Penrose is extremely accurate.} The average relative error across all bins and all applications is a mere 1.12\%. More importantly, the figure reveals a critical trend: the largest errors occur only in bins that account for a negligible fraction of the application's total execution time. Across the entire dataset, only 276 of the 19,712 bins (1.4\%) exhibited an error greater than 5\%, and the combined execution time represented by these infrequent bins amounts to just 0.064\% of the total. This result is a direct consequence of the law of large numbers; any significant deviations from sampling are confined to infrequently executed kernels that contribute minimally to the overall performance profile. This experiment provides strong evidence that our sampling methodology is robust and accurately captures the performance characteristics that matter most.

\emph{\textbf{Takeaway:} Our hardware testbed experiments validate that our simulator is accurate and that our sampling method is robust, achieving high-fidelity performance histograms with an average error of only 1.12\% while correctly capturing the behavior of dominant kernels.}

\subsection{Snippet Coverage}
We evaluated our snippet-based application identification for accuracy, latency, and storage.

\begin{table}[t]
\centering
\begin{tabular}{rrr}
\toprule
\multicolumn{1}{l}{L} & \multicolumn{1}{l}{Snippet (\%)} & \multicolumn{1}{l}{App (\%)} \\ \midrule
500 & 79.96 & 77.27 \\
1000 & 90.40 & 87.66 \\
5000 & 95.36 & 95.45 \\
10000 & 95.36 & 95.45 \\
20000 & 95.36 & 96.10 \\
\bottomrule
\end{tabular}
\smallskip
\caption{Snippet and app identification accuracy across ($L$).}\label{tab:snippetaccuracy}
\vspace{-0.4in}
\end{table}

\paragraph{Accuracy} We generated 50 snippets for each of our 154 applications and measured how accurately they were matched to their correct canonical application. Table~\ref{tab:snippetaccuracy} shows that a snippet length ($L$) of 5,000 or greater achieves over 95\% accuracy at both the snippet and application level. Shorter snippets are less distinctive and lead to more misclassifications.

\paragraph{Storage} The storage required at the AS is modest. For 2,000 applications, the MinHash table would occupy only 610 MB, which is trivial for modern servers.

\emph{\textbf{Takeaway:} With a snippet length of $L=10,000$, our application identification is over 95\% accurate, fast, and requires minimal storage, making it highly practical.}

\subsection{Overhead of AHE and User Anonymity}
While Fully Homomorphic Encryption (FHE) incurs extreme overheads, \sysname's design requires only addition, making the far more efficient Paillier Additive Homomorphic Encryption (AHE) scheme a perfect fit. The primary implementation challenges are data size expansion and computation speed. Using a 2048-bit modulus for 112-bit security, our 1KB plaintext histograms expand to 32KB of ciphertext. We now analyze the performance impact of this on clients, the aggregation server (AS), and the designer server (DS).

\begin{table}[h]
\centering
\small
\begin{tabular}{lr}
\toprule
\multicolumn{2}{c}{\textbf{Client (Latency for 1 histogram)}} \\ \midrule
Enc Speed (AMD Ryzen 7 4700U) & \multicolumn{1}{r}{431 ms} \\
Enc Speed (Intel Core i5-11500) & \multicolumn{1}{r}{105 ms} \\ \midrule
\multicolumn{2}{c}{\textbf{Aggregation Server Tput (Histograms Aggregated/sec)}} \\
Intel Xeon E-2378G & \multicolumn{1}{r}{8075} \\
Required tput for 100,000 GPUs & \multicolumn{1}{r}{33.3} \\ \midrule
\multicolumn{2}{c}{\textbf{Designer Server}} \\
Decryption Speed (Intel Xeon E-2378G) & 27 ms \\
\bottomrule
\end{tabular}
\caption{Speed measurements}\label{tab:crypto-speed}

\end{table}

Table~\ref{tab:crypto-speed} shows that the cryptographic operations are highly practical. On the client, encrypting a histogram takes 105-431 ms. Since this occurs only once every $\sim$3000 seconds, the amortized overhead is a negligible 0.0035\% to 0.014\%. On the server, a single Xeon CPU can aggregate over 8,000 histograms per second, vastly exceeding the 33.3/sec required to serve 100,000 GPUs. Decryption at the DS is also extremely fast.

\begin{figure}[tbp]
    \centering
    \includegraphics[width=\columnwidth]{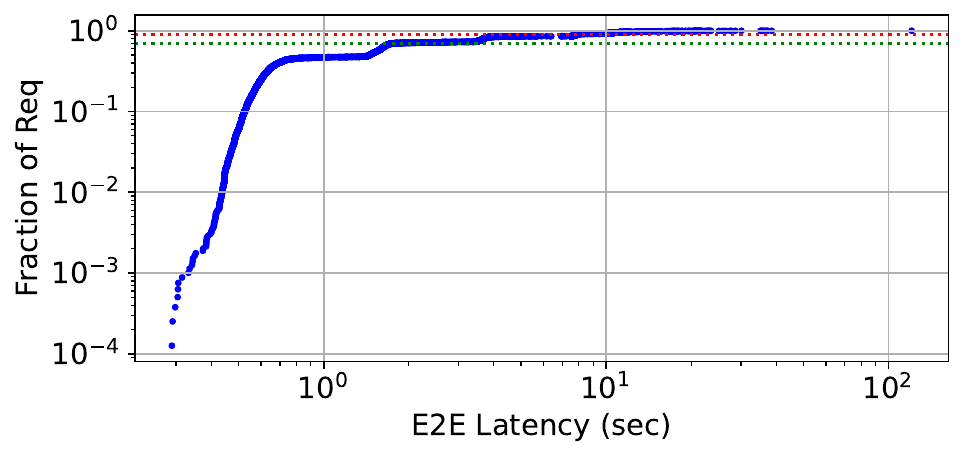}
    \caption{End-to-End Latency (log-log plot) Through Tor from Clients to AS. Green indicates 70\% of requests, red 90\%.}
    \label{fig:tore2e}
    \vspace{-0.2in}
\end{figure}

We also measured end-to-end latency through Tor by running a scaled experiment on CloudLab, simulating 10 hours of traffic. We did this measurement on a scaled system with 32 nodes sending requests every 0.96 seconds (essentially with fewer nodes we send traffic at a proportionally rate faster. 10$\times$ fewer nodes send data 10$\times$ more frequently. $0.96 sec = 3000 / (100000/32)$). Figure~\ref{fig:tore2e} shows that 70\% of requests complete in under 2 seconds and 90\% in under 8 seconds. The aggregate network traffic for 100,000 GPUs is 0.12 GB/sec, representing just 0.4\% of Tor's typical relay bandwidth~\cite{tormetrics} and well within the capacity of a standard 25 Gbps link.


\emph{\textbf{Takeaway:} Based on our measurements, \sysname's AHE and anonymity mechanisms are highly efficient, imposing minimal computation and network overheads that are easily handled by off-the-shelf hardware.}
\subsection{Cost Effectiveness and System Balance}\label{subsec:dse}
This section analyzes the system's balance and cost-effectiveness, tying together our findings on sampling, coverage, and AHE overhead. We base our analysis on a prototypical aggregation server (AS) hosted on an Equinix instance with a 25 Gbps network link and an Intel Xeon E-2378G CPU, which has an annual Total Cost of Ownership (TCO) of \$5519.

\paragraph{Computation Needs}
With our default parameters, each client transmits a partial histogram approximately every 3000 seconds. For 100,000 GPUs, this translates to a required AS processing rate of 33.3 histograms per second. Our measured throughput of 8,075 aggregations/sec vastly exceeds this requirement, confirming that AHE computation is not a bottleneck. The resulting network traffic is also well within the server's 25 Gbps link capacity.

\paragraph{Storage Growth Rate}
The use of histograms makes storage requirements minimal. For 2,000 applications, the total encrypted data per reporting period is only 64 MB, an amount easily handled by any modern server.

\paragraph{Correctness of Encrypted Computation}
We verified that our design is free from correctness issues like integer overflow. With our canonical parameters, the maximum value in any aggregated histogram bin over a 24-hour period is approximately $1.89 \times 10^{15}$. This value fits comfortably within a 64-bit integer, which our 2048-bit Paillier modulus can securely handle.

\paragraph{Scaling to Larger Histograms}
The system's costs scale linearly and gracefully with histogram size. For example, moving to 1024-bin histograms would increase client-side encryption latency to 3.3 seconds, but this still represents a negligible 0.11\% overhead when amortized over the 3000-second submission interval. AS throughput and network traffic would remain well within provisioned limits.

\emph{\textbf{Takeaway:} Our design is cost-effective and well-balanced, capable of scaling to support hundreds of thousands of GPUs with off-the-shelf hardware. The amortized system cost is low: just 6 cents per GPU per year.}

%
%

\subsection{How Can Chip Designers Use Penrose?}\label{subsec:usingpenrose}

A critical challenge for hardware architects is understanding how their designs
perform on real-world, in-the-wild workloads, which are often proprietary and
vastly more complex than public benchmarks. Penrose is designed to bridge this
exact gap. We demonstrate its utility by showing how it can provide continuous,
deep insights into complex workloads like Large Language Models (LLMs).

Consider the performance of LLMs. It is notoriously difficult to characterize,
as it depends on a complex interplay of factors like model architecture,
parallelism strategies (tensor, pipeline, expert), system scale, framework
optimizations, and dynamic runtime effects like KV cache compression. A recent
study~\cite{10.1109/SCW63240.2024.00178} required a multi-person-hour effort of
detailed, manual analysis just to characterize the performance of a few LLMs on
some inputs. Recent third-party benchmark report on a narrow
slice of traffic rapidly evolving. 

\sysname automates and scales this process. To demonstrate, we swept eight
production-scale LLMs (DeepSeekV2,GLM5,GPT-OSS-120B,Llama3-70B,Llama3-405B,Qwen3-30B,Qwen3-235B) across batch size, input context length, and output
generation length---311 configurations and \num{15.2}~million samples in
total---and reduced every run to the binned histograms \sysname would actually
transmit. No individual counter values
leave the client, all measured data.

\begin{figure}[t]
    \includegraphics[width=0.5\textwidth]{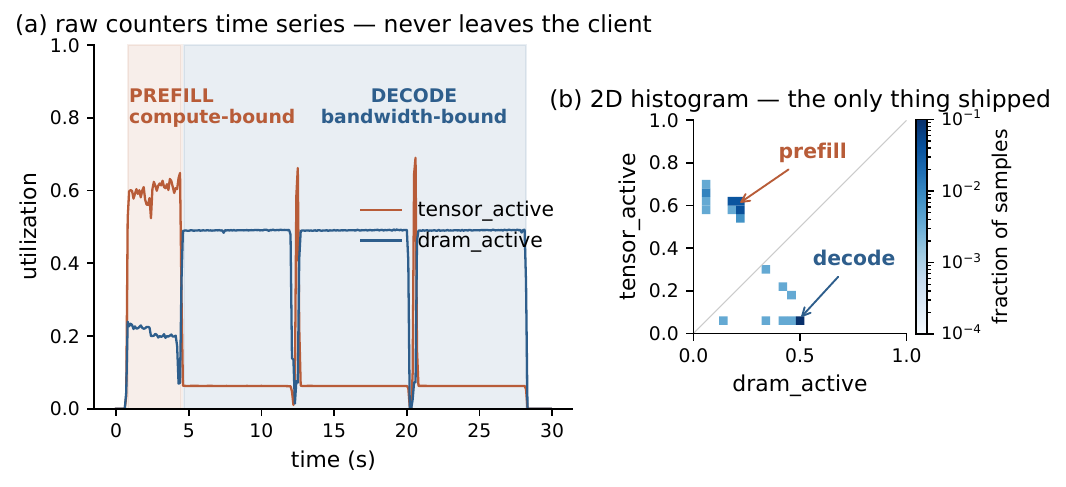}
    \caption{Time ordering is discarded; phase structure survives.
    (a)~The raw counter trace. (b)~The 2D histogram}
    \label{fig:shipped}
\end{figure}

Figure~\ref{fig:shipped} shows the class of insight this generates. Panel~(a)
is a raw trace from Llama-3-405B: the workload alternates between a prefill
phase holding \texttt{tensor\_active} $\approx 0.60$ against
\texttt{dram\_active} $\approx 0.22$, and a decode phase at $0.063$ and $0.49$
respectively---roughly a $20\times$ inversion of the tensor-to-DRAM ratio.
Panel~(b) is the shipped histogram. Because binning is order-invariant,
shuffling the trace yields an identical histogram, yet the two phases survive
as spatially separated modes. This is what makes the aggregate useful: the
run's \emph{mean} \texttt{tensor\_active} is $0.147$, which reads as a lightly
loaded accelerator and conveys nothing, while the histogram recovers a distinct
$0.60$-tensor compute mode. The joint distribution is what carries this---
marginal histograms of each counter record that a high-tensor mode and a
low-DRAM mode exist, but not that they are the same samples, understating the
prefill signature eight-fold.

This confirms, at scale and without access to raw traces, the decoupling that
motivates disaggregated prefill/decode serving: execution phases stress Tensor Cores or DRAM
bandwidth, but rarely both. It also shows how sharply it varies by
model: prefill compute peaks range from $0.11$ to $0.64$ across the eight
models we swept (Appendix~\ref{sec:per-model}), a spread no aggregate exposes.
\emph{When deployed, \sysname would track how agentic workloads are changing 
this behavior again.}


\emph{\textbf{Takeaway:} \sysname transforms a chip designer's visibility into
workload behavior, replacing sporadic, manual analysis of benchmarks with a
continuous, scalable, and privacy-preserving dashboard of real-world
application performance---recovering phase-level structure that per-counter
averages destroy, and tracking how the balance between those phases shifts as
deployments evolve.}

\mypagebreak
\vspace{-0.05in}
\section{Related Work}\label{sec:related}
\vspace{-0.05in}
\paragraph{Industry solutions} Nascent solutions include NVIDIA's GeForce Telemetry~\cite{geforce-experience} which collects some data on hardware and software configuration. Hyperscalars like Google~\cite{opentelemetry}, Meta~\cite{dynolog}, Amazon~\cite{codeguru}, Intel~\cite{intel}, and Microsoft~\cite{azure-monitor} have in-house telemetry solutions. 3rd party continuous profiling services include Datadog's continuous profiling~\cite{datadog}, Pyroscope~\cite{pyroscope}, parca~\cite{parca}, ydata-profiling,~\cite{CLEMENTE2023126585}, and Splunk's AlwaysOn profiler~\cite{splunk}. The most relevant of these is Dynolog~\cite{dynolog} --- which integrates with the PyTorch profiler and Kineto CUDA profiling library. However, to the best of our knowledge and publicly disclosed it does not capture low-level GPU performance counters which introduces issues with slowdowns. Second, it assumed knowledge/control of the application --- i.e. a chip designer (or pytorch model developer) would not be able to transparently get access to this information. These tools generally assume the entity doing the profiling has full control of the application either of a single-node or an entire cluster/data-center. 

Industry telemetry systems like DCGM and Google-Wide Profiling are designed for a single administrative domain and lack privacy guarantees. In contrast, Penrose enables privacy-preserving telemetry across organizations from opaque, user-deployed workloads. By securely identifying applications from kernel streams, it complements internal tools by giving \textbf{chip designers} unique visibility into real-world hardware performance.


\paragraph{Research profiling tools and frameworks}
 Frameworks and methodologies for profiling in the data-center include Propellor~\cite{shen2023propeller}, Accelerometer~\cite{sriraman2020accelerometer}, and Dmon~\cite{khan2021dmon} --- they don't focus on GPUs or the type of in-the-wild low-level hardware counters including the application identification challenge we solve. Large-scale application characterization studies include~\cite{davies2024journey,uta2020big,seemakhupt2023cloud,khan2021twig}. Continuous profiling was originally introduced in~\cite{anderson1997continuous}, and has since been used in a number of contexts. vprof and AutoCPA utilize continuous profiling to optimize source code~\cite{weng2023effective, rezapour2020autocpa}. \cite{safi2021distributive} implements a private continuous profiling framework for IoT-enabled networks. LightGuardian and OmniWindow propose a novel in-flight telemetry for networks like we envision for compute~\cite{265015,sun2023omniwindow}. Basic block vectors were proposed as way for program characterization~and later work expanded for automatically characterizing large scale program behavior and simulation strategies, bearing similarity to our snippet classifcation~\cite{sherwood2001basic,sherwood2002automatically,lau2004structures,sherwood2003discovering,hamerly2005simpoint,lau2006selecting,sherwood2003phase}.

\paragraph{Privacy}
Privacy-Preserving Machine Learning (PPML) encompasses secure training, inference, and federated learning~\cite{PPML-msft,xu2021PPML,secureml}. Our focus is on computing a function $f$ over private datasets, without revealing the data—specifically, via secure histogram aggregation. This is a simplified instance of broader secure aggregation problems~\cite{secure-agg}. Relevant applications of AHE and FHE  include~\cite{irene,fhesurvey,cryptdb,damgard-jurik,nikolaenko,irene-ppml}.



\section{Conclusion}\label{sec:conc}
As the complexity of both applications and hardware increases, GPU chip manufacturers face a significant question: how to gather comprehensive performance characteristics and value profiles from GPUs deployed in real-world scenarios? Such data, encompassing the types of kernels executed and the time spent in each, is crucial for optimizing chip design and enhancing application performance This paper undertakes an ambitious system to provide extensive data in a privacy-preserving fashion combining ideas of sampling, data reduction through histograms, and AHE-based computation to preserve privacy. Our system \sysname shows that a single aggregation server matching a typical \textit{small} compute instance cloud server can manage and collect profiles for 100,000 GPUs. At typical load rate of 10\% activity and with 100,000 participating GPUs, \sysname achieves 99\% coverage of applications in one day, thus being able to provide a type of streaming data of application behavior change.


\bibliographystyle{ACM-Reference-Format}
\bibliography{references,fhe,ahe}


\begin{thebibliography}{85}


\ifx \showCODEN    \undefined \def \showCODEN     #1{\unskip}     \fi
\ifx \showISBNx    \undefined \def \showISBNx     #1{\unskip}     \fi
\ifx \showISBNxiii \undefined \def \showISBNxiii  #1{\unskip}     \fi
\ifx \showISSN     \undefined \def \showISSN      #1{\unskip}     \fi
\ifx \showLCCN     \undefined \def \showLCCN      #1{\unskip}     \fi
\ifx \shownote     \undefined \def \shownote      #1{#1}          \fi
\ifx \showarticletitle \undefined \def \showarticletitle #1{#1}   \fi
\ifx \showURL      \undefined \def \showURL       {\relax}        \fi
\providecommand\bibfield[2]{#2}
\providecommand\bibinfo[2]{#2}
\providecommand\natexlab[1]{#1}
\providecommand\showeprint[2][]{arXiv:#2}

\bibitem[ACO(2023)]%
        {ACORN}
 \bibinfo{year}{2023}\natexlab{}.
\newblock \showarticletitle{ACORN: input validation for secure aggregation}. In \bibinfo{booktitle}{\emph{32nd USENIX Security Symposium (USENIX Security 23)}}.
\newblock


\bibitem[Acar et~al\mbox{.}(2018)]%
        {10.1145/3214303}
\bibfield{author}{\bibinfo{person}{Abbas Acar}, \bibinfo{person}{Hidayet Aksu}, \bibinfo{person}{A.~Selcuk Uluagac}, {and} \bibinfo{person}{Mauro Conti}.} \bibinfo{year}{2018}\natexlab{}.
\newblock \showarticletitle{A Survey on Homomorphic Encryption Schemes: Theory and Implementation}.
\newblock \bibinfo{journal}{\emph{ACM Comput. Surv.}} \bibinfo{volume}{51}, \bibinfo{number}{4}, Article \bibinfo{articleno}{79} (\bibinfo{date}{July} \bibinfo{year}{2018}), \bibinfo{numpages}{35}~pages.
\newblock
\showISSN{0360-0300}
\href{https://doi.org/10.1145/3214303}{doi:\nolinkurl{10.1145/3214303}}


\bibitem[Amazon({[n.\,d.]})]%
        {codeguru}
\bibfield{author}{\bibinfo{person}{Amazon}.} \bibinfo{year}{[n.\,d.]}\natexlab{}.
\newblock \bibinfo{title}{CodeGuru. \url{https://aws.amazon.com/blogs/machine-learning/optimizing-application-performance-with-amazon-codeguru-profiler/}}.
\newblock


\bibitem[Anderson et~al\mbox{.}(1997)]%
        {anderson1997continuous}
\bibfield{author}{\bibinfo{person}{Jennifer~M Anderson}, \bibinfo{person}{Lance~M Berc}, \bibinfo{person}{Jeffrey Dean}, \bibinfo{person}{Sanjay Ghemawat}, \bibinfo{person}{Monika~R Henzinger}, \bibinfo{person}{Shun-Tak~A Leung}, \bibinfo{person}{Richard~L Sites}, \bibinfo{person}{Mark~T Vandevoorde}, \bibinfo{person}{Carl~A Waldspurger}, {and} \bibinfo{person}{William~E Weihl}.} \bibinfo{year}{1997}\natexlab{}.
\newblock \showarticletitle{Continuous profiling: Where have all the cycles gone?}
\newblock \bibinfo{journal}{\emph{ACM Transactions on Computer Systems (TOCS)}} \bibinfo{volume}{15}, \bibinfo{number}{4} (\bibinfo{year}{1997}), \bibinfo{pages}{357--390}.
\newblock


\bibitem[Apsey et~al\mbox{.}(2023)]%
        {h100-attestation}
\bibfield{author}{\bibinfo{person}{Emily Apsey}, \bibinfo{person}{Phil Rogers}, \bibinfo{person}{Michael O'Connor}, {and} \bibinfo{person}{Rob Nertney}.} \bibinfo{year}{2023}\natexlab{}.
\newblock \bibinfo{title}{Confidential Computing on NVIDIA H100 GPUs for Secure and Trustworthy AI. NVIDIA Techblog \url{https://developer.nvidia.com/blog/confidential-computing-on-h100-gpus-for-secure-and-trustworthy-ai/}}.
\newblock


\bibitem[Barker(2020)]%
        {nist}
\bibfield{author}{\bibinfo{person}{Elaine Barker}.} \bibinfo{year}{2020}\natexlab{}.
\newblock \bibinfo{title}{{NIST} Special Publication 800-57 Part 1 Revision 5. Recommendation for Key Management}.
\newblock \bibinfo{howpublished}{\url{https://doi.org/10.6028/NIST.SP.800-57pt1r5}}.
\newblock


\bibitem[Bonawitz et~al\mbox{.}(2016)]%
        {secure-agg}
\bibfield{author}{\bibinfo{person}{K.~A. Bonawitz}, \bibinfo{person}{Vladimir Ivanov}, \bibinfo{person}{Ben Kreuter}, \bibinfo{person}{Antonio Marcedone}, \bibinfo{person}{H.~Brendan McMahan}, \bibinfo{person}{Sarvar Patel}, \bibinfo{person}{Daniel Ramage}, \bibinfo{person}{Aaron Segal}, {and} \bibinfo{person}{Karn Seth}.} \bibinfo{year}{2016}\natexlab{}.
\newblock \showarticletitle{Practical Secure Aggregation for Federated Learning on User-Held Data}. In \bibinfo{booktitle}{\emph{NeuRIPS Workshop on Private Multi-Party Machine Learning}}.
\newblock
\urldef\tempurl%
\url{https://arxiv.org/abs/1611.04482}
\showURL{%
\tempurl}


\bibitem[Boneh and Shoup(2023)]%
        {boneh-shoup}
\bibfield{author}{\bibinfo{person}{Dan Boneh} {and} \bibinfo{person}{Victor Shoup}.} \bibinfo{year}{2023}\natexlab{}.
\newblock \bibinfo{title}{A Graduate Course in Applied Cryptography}.
\newblock
\urldef\tempurl%
\url{https://toc.cryptobook.us/}
\showURL{%
\tempurl}


\bibitem[Brickell et~al\mbox{.}(2004)]%
        {brickell2004direct}
\bibfield{author}{\bibinfo{person}{Ernie Brickell}, \bibinfo{person}{Jan Camenisch}, {and} \bibinfo{person}{Liqun Chen}.} \bibinfo{year}{2004}\natexlab{}.
\newblock \showarticletitle{Direct anonymous attestation}. In \bibinfo{booktitle}{\emph{Proceedings of the 11th ACM conference on Computer and communications security}}. \bibinfo{pages}{132--145}.
\newblock


\bibitem[Brickell et~al\mbox{.}(2008)]%
        {brickell2008new}
\bibfield{author}{\bibinfo{person}{Ernie Brickell}, \bibinfo{person}{Liqun Chen}, {and} \bibinfo{person}{Jiangtao Li}.} \bibinfo{year}{2008}\natexlab{}.
\newblock \showarticletitle{A new direct anonymous attestation scheme from bilinear maps}. In \bibinfo{booktitle}{\emph{Trusted Computing-Challenges and Applications: First International Conference on Trusted Computing and Trust in Information Technologies, Trust 2008 Villach, Austria, March 11-12, 2008 Proceedings 1}}. Springer, \bibinfo{pages}{166--178}.
\newblock


\bibitem[Brickell and Li(2007)]%
        {brickell2007enhanced}
\bibfield{author}{\bibinfo{person}{Ernie Brickell} {and} \bibinfo{person}{Jiangtao Li}.} \bibinfo{year}{2007}\natexlab{}.
\newblock \showarticletitle{Enhanced privacy ID: A direct anonymous attestation scheme with enhanced revocation capabilities}. In \bibinfo{booktitle}{\emph{Proceedings of the 2007 ACM workshop on Privacy in electronic society}}. \bibinfo{pages}{21--30}.
\newblock


\bibitem[Camenisch et~al\mbox{.}(2016)]%
        {camenisch2016anonymous}
\bibfield{author}{\bibinfo{person}{Jan Camenisch}, \bibinfo{person}{Manu Drijvers}, {and} \bibinfo{person}{Anja Lehmann}.} \bibinfo{year}{2016}\natexlab{}.
\newblock \showarticletitle{Anonymous attestation using the strong diffie hellman assumption revisited}. In \bibinfo{booktitle}{\emph{Trust and Trustworthy Computing: 9th International Conference, TRUST 2016, Vienna, Austria, August 29-30, 2016, Proceedings 9}}. Springer, \bibinfo{pages}{1--20}.
\newblock


\bibitem[Chaum(1981)]%
        {chaum1981untraceable}
\bibfield{author}{\bibinfo{person}{David~L Chaum}.} \bibinfo{year}{1981}\natexlab{}.
\newblock \showarticletitle{Untraceable electronic mail, return addresses, and digital pseudonyms}.
\newblock \bibinfo{journal}{\emph{Commun. ACM}} \bibinfo{volume}{24}, \bibinfo{number}{2} (\bibinfo{year}{1981}), \bibinfo{pages}{84--90}.
\newblock


\bibitem[Chitty-Venkata et~al\mbox{.}(2025)]%
        {10.1109/SCW63240.2024.00178}
\bibfield{author}{\bibinfo{person}{Krishna~Teja Chitty-Venkata}, \bibinfo{person}{Siddhisanket Raskar}, \bibinfo{person}{Bharat Kale}, \bibinfo{person}{Farah Ferdaus}, \bibinfo{person}{Aditya Tanikanti}, \bibinfo{person}{Ken Raffenetti}, \bibinfo{person}{Valerie Taylor}, \bibinfo{person}{Murali Emani}, {and} \bibinfo{person}{Venkatram Vishwanath}.} \bibinfo{year}{2025}\natexlab{}.
\newblock \showarticletitle{LLM-Inference-Bench: Inference Benchmarking of Large Language Models on AI Accelerators}. In \bibinfo{booktitle}{\emph{Proceedings of the SC '24 Workshops of the International Conference on High Performance Computing, Network, Storage, and Analysis}} (Atlanta, GA, USA) \emph{(\bibinfo{series}{SC-W '24})}. \bibinfo{publisher}{IEEE Press}, \bibinfo{pages}{1362–1379}.
\newblock
\showISBNx{9798350355543}
\href{https://doi.org/10.1109/SCW63240.2024.00178}{doi:\nolinkurl{10.1109/SCW63240.2024.00178}}


\bibitem[Choquette(2022)]%
        {choquette2022nvidia}
\bibfield{author}{\bibinfo{person}{Jack Choquette}.} \bibinfo{year}{2022}\natexlab{}.
\newblock \showarticletitle{Nvidia hopper gpu: Scaling performance}. In \bibinfo{booktitle}{\emph{2022 IEEE Hot Chips 34 Symposium (HCS)}}. IEEE Computer Society, \bibinfo{pages}{1--46}.
\newblock


\bibitem[Chowdhury et~al\mbox{.}(2022)]%
        {Eiffel}
\bibfield{author}{\bibinfo{person}{Amrita~Roy Chowdhury}, \bibinfo{person}{Chuan Guo}, \bibinfo{person}{Somesh Jha}, {and} \bibinfo{person}{Laurens van~der Maaten}.} \bibinfo{year}{2022}\natexlab{}.
\newblock \bibinfo{title}{EIFFeL: Ensuring Integrity for Federated Learning}.
\newblock
\showeprint[arxiv]{2112.12727}~[cs.CR]
\urldef\tempurl%
\url{https://arxiv.org/abs/2112.12727}
\showURL{%
\tempurl}


\bibitem[Clemente et~al\mbox{.}(2023)]%
        {CLEMENTE2023126585}
\bibfield{author}{\bibinfo{person}{Fabiana Clemente}, \bibinfo{person}{Gonçalo~Martins Ribeiro}, \bibinfo{person}{Alexandre Quemy}, \bibinfo{person}{Miriam~Seoane Santos}, \bibinfo{person}{Ricardo~Cardoso Pereira}, {and} \bibinfo{person}{Alex Barros}.} \bibinfo{year}{2023}\natexlab{}.
\newblock \showarticletitle{ydata-profiling: Accelerating data-centric AI with high-quality data}.
\newblock \bibinfo{journal}{\emph{Neurocomputing}}  \bibinfo{volume}{554} (\bibinfo{year}{2023}), \bibinfo{pages}{126585}.
\newblock
\showISSN{0925-2312}
\href{https://doi.org/10.1016/j.neucom.2023.126585}{doi:\nolinkurl{10.1016/j.neucom.2023.126585}}


\bibitem[Corporation(2025)]%
        {nvidia_nsight_systems}
\bibfield{author}{\bibinfo{person}{NVIDIA Corporation}.} \bibinfo{year}{2025}\natexlab{}.
\newblock \bibinfo{title}{Nsight Systems - System-Wide Performance Analysis Tool}.
\newblock
\shownote{Accessed: April 10, 2025}.
\newblock
\urldef\tempurl%
\url{https://developer.nvidia.com/nsight-systems}
\showURL{%
\tempurl}


\bibitem[Coutinho({[n.\,d.]})]%
        {dynolog}
\bibfield{author}{\bibinfo{person}{ByBrian Coutinho}.} \bibinfo{year}{[n.\,d.]}\natexlab{}.
\newblock \bibinfo{title}{Dynolog: Open source system observability. \url{https://developers.facebook.com/blog/post/2022/11/16/dynolog-open-source-system-observability/}}.
\newblock
\urldef\tempurl%
\url{https://developers.facebook.com/blog/post/2022/11/16/dynolog-open-source-system-observability/}
\showURL{%
\tempurl}


\bibitem[Dally et~al\mbox{.}(2021)]%
        {9623445}
\bibfield{author}{\bibinfo{person}{William~J. Dally}, \bibinfo{person}{Stephen~W. Keckler}, {and} \bibinfo{person}{David~B. Kirk}.} \bibinfo{year}{2021}\natexlab{}.
\newblock \showarticletitle{Evolution of the Graphics Processing Unit (GPU)}.
\newblock \bibinfo{journal}{\emph{IEEE Micro}} \bibinfo{volume}{41}, \bibinfo{number}{6} (\bibinfo{year}{2021}), \bibinfo{pages}{42--51}.
\newblock
\href{https://doi.org/10.1109/MM.2021.3113475}{doi:\nolinkurl{10.1109/MM.2021.3113475}}


\bibitem[Damgard et~al\mbox{.}(2010)]%
        {damgard-jurik}
\bibfield{author}{\bibinfo{person}{I. Damgard}, \bibinfo{person}{M. Jurik}, {and} \bibinfo{person}{J.B. Nielsen}.} \bibinfo{year}{2010}\natexlab{}.
\newblock \showarticletitle{A generalisation of Paillier’s public-key system with applications to electronic voting}.
\newblock \bibinfo{journal}{\emph{Int. J. Inf. Sec.}} (\bibinfo{year}{2010}).
\newblock


\bibitem[Datadog({[n.\,d.]})]%
        {datadog}
\bibfield{author}{\bibinfo{person}{Datadog}.} \bibinfo{year}{[n.\,d.]}\natexlab{}.
\newblock \bibinfo{title}{Datadog Continuous Profiler. \url{https://www.datadoghq.com/product/code-profiling/}}.
\newblock


\bibitem[Davies et~al\mbox{.}(2024)]%
        {davies2024journey}
\bibfield{author}{\bibinfo{person}{Michael Davies}, \bibinfo{person}{Ian McDougall}, \bibinfo{person}{Selvaraj Anandaraj}, \bibinfo{person}{Deep Machchhar}, \bibinfo{person}{Rithik Jain}, {and} \bibinfo{person}{Karthikeyan Sankaralingam}.} \bibinfo{year}{2024}\natexlab{}.
\newblock \showarticletitle{A Journey of a 1,000 Kernels Begins with a Single Step: A Retrospective of Deep Learning on GPUs}. In \bibinfo{booktitle}{\emph{Proceedings of the 29th ACM International Conference on Architectural Support for Programming Languages and Operating Systems, Volume 2}}. \bibinfo{pages}{20--36}.
\newblock


\bibitem[Dingledine et~al\mbox{.}(2004)]%
        {dingledine2004tor}
\bibfield{author}{\bibinfo{person}{Roger Dingledine}, \bibinfo{person}{Nick Mathewson}, \bibinfo{person}{Paul~F Syverson}, {et~al\mbox{.}}} \bibinfo{year}{2004}\natexlab{}.
\newblock \showarticletitle{Tor: The second-generation onion router.}. In \bibinfo{booktitle}{\emph{USENIX security symposium}}, Vol.~\bibinfo{volume}{4}. \bibinfo{pages}{303--320}.
\newblock


\bibitem[Equinix({[n.\,d.]})]%
        {equinix}
\bibfield{author}{\bibinfo{person}{Equinix}.} \bibinfo{year}{[n.\,d.]}\natexlab{}.
\newblock \bibinfo{title}{m3.small.x86. Specifications \& Pricing \url{https://deploy.equinix.com/product/servers/m3-small/}}.
\newblock


\bibitem[Erlingsson et~al\mbox{.}(2014)]%
        {ldp}
\bibfield{author}{\bibinfo{person}{Ulfar Erlingsson}, \bibinfo{person}{Vasyl Pihur}, {and} \bibinfo{person}{Aleksandra Korolova}.} \bibinfo{year}{2014}\natexlab{}.
\newblock \showarticletitle{RAPPOR: Randomized Aggregatable Privacy-Preserving Ordinal Response}. In \bibinfo{booktitle}{\emph{CCS}}.
\newblock


\bibitem[Giacomelli et~al\mbox{.}(2017)]%
        {irene}
\bibfield{author}{\bibinfo{person}{Irene Giacomelli}, \bibinfo{person}{Somesh Jha}, \bibinfo{person}{Marc Joye}, \bibinfo{person}{C.~David Page}, {and} \bibinfo{person}{Kyonghwan Yoon}.} \bibinfo{year}{2017}\natexlab{}.
\newblock \bibinfo{title}{Privacy-Preserving Ridge Regression with only Linearly-Homomorphic Encryption}.
\newblock \bibinfo{howpublished}{Cryptology ePrint Archive, Paper 2017/979}.
\newblock
\shownote{\url{https://eprint.iacr.org/2017/979}}.
\newblock
\urldef\tempurl%
\url{https://eprint.iacr.org/2017/979}
\showURL{%
\tempurl}


\bibitem[Giacomelli et~al\mbox{.}(2018)]%
        {irene-ppml}
\bibfield{author}{\bibinfo{person}{Irene Giacomelli}, \bibinfo{person}{Somesh Jha}, \bibinfo{person}{Marc Joye}, \bibinfo{person}{C~David Page}, {and} \bibinfo{person}{Kyonghwan Yoon}.} \bibinfo{year}{2018}\natexlab{}.
\newblock \showarticletitle{Privacy-preserving ridge regression with only linearly-homomorphic encryption}. In \bibinfo{booktitle}{\emph{Applied Cryptography and Network Security (ACNS)}}.
\newblock


\bibitem[Google({[n.\,d.]})]%
        {opentelemetry}
Google \bibinfo{year}{[n.\,d.]}\natexlab{}.
\newblock \bibinfo{title}{Open Telemetry. \url{https://cloud.google.com/learn/what-is-opentelemetry}}.
\newblock


\bibitem[Grafana({[n.\,d.]})]%
        {pyroscope}
\bibfield{author}{\bibinfo{person}{Grafana}.} \bibinfo{year}{[n.\,d.]}\natexlab{}.
\newblock \bibinfo{title}{Pyroscope. \url{https://github.com/grafana/pyroscope}}.
\newblock


\bibitem[Grubbs et~al\mbox{.}(2019)]%
        {grubbs}
\bibfield{author}{\bibinfo{person}{Paul Grubbs}, \bibinfo{person}{Marie-Sarah Lacharite}, \bibinfo{person}{Brice Minaud}, {and} \bibinfo{person}{Kenneth~G Paterson}.} \bibinfo{year}{2019}\natexlab{}.
\newblock \showarticletitle{Learning to reconstruct: Statistical learning theory and encrypted database attacks}. In \bibinfo{booktitle}{\emph{IEEE Symposium on Security and Privacy (SP)}}.
\newblock


\bibitem[Hamerly et~al\mbox{.}(2005)]%
        {hamerly2005simpoint}
\bibfield{author}{\bibinfo{person}{Greg Hamerly}, \bibinfo{person}{Erez Perelman}, \bibinfo{person}{Jeremy Lau}, {and} \bibinfo{person}{Brad Calder}.} \bibinfo{year}{2005}\natexlab{}.
\newblock \showarticletitle{Simpoint 3.0: Faster and more flexible program phase analysis}.
\newblock \bibinfo{journal}{\emph{Journal of Instruction Level Parallelism}} \bibinfo{volume}{7}, \bibinfo{number}{4} (\bibinfo{year}{2005}), \bibinfo{pages}{1--28}.
\newblock


\bibitem[Hao et~al\mbox{.}(2023)]%
        {hao2023torchbench}
\bibfield{author}{\bibinfo{person}{Yueming Hao}, \bibinfo{person}{Xu Zhao}, \bibinfo{person}{Bin Bao}, \bibinfo{person}{David Berard}, \bibinfo{person}{Will Constable}, \bibinfo{person}{Adnan Aziz}, {and} \bibinfo{person}{Xu Liu}.} \bibinfo{year}{2023}\natexlab{}.
\newblock \showarticletitle{Torchbench: Benchmarking pytorch with high api surface coverage}.
\newblock \bibinfo{journal}{\emph{arXiv preprint arXiv:2304.14226}} (\bibinfo{year}{2023}).
\newblock


\bibitem[Hranick{\`y} et~al\mbox{.}(2019)]%
        {hranicky2019distributed}
\bibfield{author}{\bibinfo{person}{Radek Hranick{\`y}}, \bibinfo{person}{Luk{\'a}{\v{s}} Zobal}, \bibinfo{person}{Ond{\v{r}}ej Ry{\v{s}}av{\`y}}, {and} \bibinfo{person}{Du{\v{s}}an Kol{\'a}{\v{r}}}.} \bibinfo{year}{2019}\natexlab{}.
\newblock \showarticletitle{Distributed password cracking with BOINC and hashcat}.
\newblock \bibinfo{journal}{\emph{Digital Investigation}}  \bibinfo{volume}{30} (\bibinfo{year}{2019}), \bibinfo{pages}{161--172}.
\newblock


\bibitem[Inc.(2023)]%
        {apple2023icloud}
\bibfield{author}{\bibinfo{person}{Apple Inc.}} \bibinfo{year}{2023}\natexlab{}.
\newblock \bibinfo{title}{About iCloud Private Relay}.
\newblock
\shownote{Accessed: April 10, 2025}.
\newblock
\urldef\tempurl%
\url{https://support.apple.com/en-us/102602}
\showURL{%
\tempurl}


\bibitem[Intel(2026a)]%
        {intel}
\bibfield{author}{\bibinfo{person}{Intel}.} \bibinfo{year}{2026}\natexlab{a}.
\newblock \bibinfo{title}{Intel Continuous Profiler. \url{https://www.intc.com/news-events/press-releases/detail/1683/intel-releases-continuous-profiler-to-increase-cpu}}.
\newblock
\shownote{Accessed: September 22, 2026}.
\newblock


\bibitem[Intel(2026b)]%
        {intel-paillier}
\bibfield{author}{\bibinfo{person}{Intel}.} \bibinfo{year}{2026}\natexlab{b}.
\newblock \bibinfo{title}{Intel Paillier Cryptosystem Library. \url{https://github.com/intel/pailliercryptolib/tree/development}}.
\newblock
\shownote{Accessed: September 22, 2026}.
\newblock


\bibitem[Intel(2026c)]%
        {intel-pmt}
\bibfield{author}{\bibinfo{person}{Intel}.} \bibinfo{year}{2026}\natexlab{c}.
\newblock \bibinfo{title}{Intel PMT. \url{https://github.com/intel/Intel-PMT}}.
\newblock
\shownote{Accessed: September 22, 2026}.
\newblock


\bibitem[Johnson et~al\mbox{.}(2016)]%
        {johnson2016intel}
\bibfield{author}{\bibinfo{person}{Simon Johnson}, \bibinfo{person}{Vinnie Scarlata}, \bibinfo{person}{Carlos Rozas}, \bibinfo{person}{Ernie Brickell}, \bibinfo{person}{Frank Mckeen}, {et~al\mbox{.}}} \bibinfo{year}{2016}\natexlab{}.
\newblock \showarticletitle{Intel software guard extensions: EPID provisioning and attestation services}.
\newblock \bibinfo{journal}{\emph{White Paper}} \bibinfo{volume}{1}, \bibinfo{number}{1-10} (\bibinfo{year}{2016}), \bibinfo{pages}{119}.
\newblock


\bibitem[Katz and Lindell(2020)]%
        {Katz-Lindell}
\bibfield{author}{\bibinfo{person}{Jonathan Katz} {and} \bibinfo{person}{Yehuda Lindell}.} \bibinfo{year}{2020}\natexlab{}.
\newblock \bibinfo{booktitle}{\emph{Introduction to Modern Cryptography: Third Edition}}.
\newblock \bibinfo{publisher}{Chapman \& Hall/CRC Cryptography and Network Security Series) 3rd Edition}.
\newblock


\bibitem[Khan et~al\mbox{.}(2021a)]%
        {khan2021twig}
\bibfield{author}{\bibinfo{person}{Tanvir~Ahmed Khan}, \bibinfo{person}{Nathan Brown}, \bibinfo{person}{Akshitha Sriraman}, \bibinfo{person}{Niranjan~K Soundararajan}, \bibinfo{person}{Rakesh Kumar}, \bibinfo{person}{Joseph Devietti}, \bibinfo{person}{Sreenivas Subramoney}, \bibinfo{person}{Gilles~A Pokam}, \bibinfo{person}{Heiner Litz}, {and} \bibinfo{person}{Baris Kasikci}.} \bibinfo{year}{2021}\natexlab{a}.
\newblock \showarticletitle{Twig: Profile-guided btb prefetching for data center applications}. In \bibinfo{booktitle}{\emph{MICRO-54: 54th Annual IEEE/ACM International Symposium on Microarchitecture}}. \bibinfo{pages}{816--829}.
\newblock


\bibitem[Khan et~al\mbox{.}(2021b)]%
        {khan2021dmon}
\bibfield{author}{\bibinfo{person}{Tanvir~Ahmed Khan}, \bibinfo{person}{Ian Neal}, \bibinfo{person}{Gilles Pokam}, \bibinfo{person}{Barzan Mozafari}, {and} \bibinfo{person}{Baris Kasikci}.} \bibinfo{year}{2021}\natexlab{b}.
\newblock \showarticletitle{Dmon: Efficient detection and correction of data locality problems using selective profiling}. In \bibinfo{booktitle}{\emph{15th $\{$USENIX$\}$ Symposium on Operating Systems Design and Implementation ($\{$OSDI$\}$ 21)}}. \bibinfo{pages}{163--181}.
\newblock


\bibitem[Lau et~al\mbox{.}(2006)]%
        {lau2006selecting}
\bibfield{author}{\bibinfo{person}{Jeremy Lau}, \bibinfo{person}{Erez Perelman}, {and} \bibinfo{person}{Brad Calder}.} \bibinfo{year}{2006}\natexlab{}.
\newblock \showarticletitle{Selecting software phase markers with code structure analysis}. In \bibinfo{booktitle}{\emph{International Symposium on Code Generation and Optimization (CGO'06)}}. IEEE, \bibinfo{pages}{12--pp}.
\newblock


\bibitem[Lau et~al\mbox{.}(2004)]%
        {lau2004structures}
\bibfield{author}{\bibinfo{person}{Jeremy Lau}, \bibinfo{person}{Stefan Schoemackers}, {and} \bibinfo{person}{Brad Calder}.} \bibinfo{year}{2004}\natexlab{}.
\newblock \showarticletitle{Structures for phase classification}. In \bibinfo{booktitle}{\emph{IEEE International Symposium on-ISPASS Performance Analysis of Systems and Software, 2004}}. IEEE, \bibinfo{pages}{57--67}.
\newblock


\bibitem[Leith(2021)]%
        {leith2021mobile}
\bibfield{author}{\bibinfo{person}{Douglas~J Leith}.} \bibinfo{year}{2021}\natexlab{}.
\newblock \showarticletitle{Mobile handset privacy: Measuring the data ios and android send to apple and google}. In \bibinfo{booktitle}{\emph{International Conference on Security and Privacy in Communication Systems}}. Springer, \bibinfo{pages}{231--251}.
\newblock


\bibitem[Levine et~al\mbox{.}(2006)]%
        {levine2006survey}
\bibfield{author}{\bibinfo{person}{Brian~Neil Levine}, \bibinfo{person}{Clay Shields}, {and} \bibinfo{person}{N~Boris Margolin}.} \bibinfo{year}{2006}\natexlab{}.
\newblock \showarticletitle{A survey of solutions to the sybil attack}.
\newblock \bibinfo{journal}{\emph{University of Massachusetts Amherst, Amherst, MA}}  \bibinfo{volume}{7} (\bibinfo{year}{2006}), \bibinfo{pages}{224}.
\newblock


\bibitem[Lubeck(2013)]%
        {lubeck2013johntheripper}
\bibfield{author}{\bibinfo{person}{Tyler Lubeck}.} \bibinfo{year}{2013}\natexlab{}.
\newblock \bibinfo{booktitle}{\emph{Distributed Password Cracking with John the Ripper}}.
\newblock \bibinfo{type}{{T}echnical {R}eport}. \bibinfo{institution}{Tufts University}.
\newblock
\urldef\tempurl%
\url{https://www.cs.tufts.edu/comp/116/archive/fall2013/tlubeck.pdf}
\showURL{%
\tempurl}


\bibitem[Marcolla et~al\mbox{.}(2022)]%
        {fhesurvey}
\bibfield{author}{\bibinfo{person}{Chiara Marcolla}, \bibinfo{person}{Victor Sucasas}, \bibinfo{person}{Marc Manzano}, \bibinfo{person}{Riccardo Bassoli}, \bibinfo{person}{Frank~H.P. Fitzek}, {and} \bibinfo{person}{Najwa Aaraj}.} \bibinfo{year}{2022}\natexlab{}.
\newblock \bibinfo{title}{Survey on Fully Homomorphic Encryption, Theory, and Applications}.
\newblock \bibinfo{howpublished}{Cryptology {ePrint} Archive, Paper 2022/1602}.
\newblock
\href{https://doi.org/10.1109/JPROC.2022.3205665}{doi:\nolinkurl{10.1109/JPROC.2022.3205665}}


\bibitem[meta({[n.\,d.]})]%
        {torchbench2}
meta \bibinfo{year}{[n.\,d.]}\natexlab{}.
\newblock \bibinfo{title}{Will Constable et al. 2020. TorchBench: a collection of open source benchmarks for PyTorch performance and usability evaluation. \url{http s://github.com/pytorch/benchmark.}}
\newblock


\bibitem[Microsoft({[n.\,d.]})]%
        {azure-monitor}
\bibfield{author}{\bibinfo{person}{Microsoft}.} \bibinfo{year}{[n.\,d.]}\natexlab{}.
\newblock \bibinfo{title}{Azure Monitor. \url{https://learn.microsoft.com/en-us/azure/azure-monitor/getting-started}}.
\newblock


\bibitem[Mittal and Vaishay(2019)]%
        {mittal2019survey}
\bibfield{author}{\bibinfo{person}{Sparsh Mittal} {and} \bibinfo{person}{Shraiysh Vaishay}.} \bibinfo{year}{2019}\natexlab{}.
\newblock \showarticletitle{A survey of techniques for optimizing deep learning on GPUs}.
\newblock \bibinfo{journal}{\emph{Journal of Systems Architecture}}  \bibinfo{volume}{99} (\bibinfo{year}{2019}), \bibinfo{pages}{101635}.
\newblock


\bibitem[Mohassel and Zhang(2017)]%
        {secureml}
\bibfield{author}{\bibinfo{person}{Payman Mohassel} {and} \bibinfo{person}{Yupeng Zhang}.} \bibinfo{year}{2017}\natexlab{}.
\newblock \showarticletitle{SecureML: A System for Scalable Privacy-Preserving Machine Learning}. In \bibinfo{booktitle}{\emph{2017 IEEE Symposium on Security and Privacy (SP)}}. \bibinfo{pages}{19--38}.
\newblock
\href{https://doi.org/10.1109/SP.2017.12}{doi:\nolinkurl{10.1109/SP.2017.12}}


\bibitem[Narayanan and Shmatikov(2008)]%
        {shmatikov}
\bibfield{author}{\bibinfo{person}{Arvind Narayanan} {and} \bibinfo{person}{Vitaly Shmatikov}.} \bibinfo{year}{2008}\natexlab{}.
\newblock \showarticletitle{Robust de-anonymization of large sparse datasets}. In \bibinfo{booktitle}{\emph{IEEE Symposium on Security and Privacy}}.
\newblock


\bibitem[Nikolaenko et~al\mbox{.}(2013)]%
        {nikolaenko}
\bibfield{author}{\bibinfo{person}{V. Nikolaenko}, \bibinfo{person}{U. Weinsberg}, \bibinfo{person}{S. Ioannidis}, \bibinfo{person}{M. Joye}, \bibinfo{person}{D. Boneh}, {and} \bibinfo{person}{N. Taft}.} \bibinfo{year}{2013}\natexlab{}.
\newblock \showarticletitle{Privacy-preserving ridge regression on hundreds of millions of records}. In \bibinfo{booktitle}{\emph{IEEE Symposium on Security and Privacy}}.
\newblock


\bibitem[NVIDIA({[n.\,d.]})]%
        {geforce-experience}
\bibfield{author}{\bibinfo{person}{NVIDIA}.} \bibinfo{year}{[n.\,d.]}\natexlab{}.
\newblock \bibinfo{title}{GEFORCE EXperience. \url{https://www.nvidia.com/en-us/geforce/geforce-experience/}}.
\newblock


\bibitem[nvidia({[n.\,d.]})]%
        {cudakernels}
nvidia \bibinfo{year}{[n.\,d.]}\natexlab{}.
\newblock \bibinfo{title}{GPU Accelerated Libraries. \url{https://developer.nvidia.com/gpu-accelerated-libraries}}.
\newblock


\bibitem[NVIDIA(2026)]%
        {ncu-counters}
\bibfield{author}{\bibinfo{person}{NVIDIA}.} \bibinfo{year}{2026}\natexlab{}.
\newblock \bibinfo{title}{Kernel Profiling Guide}.
\newblock
\shownote{Accessed: September 22, 2026}.
\newblock
\urldef\tempurl%
\url{https://docs.nvidia.com/nsight-compute/ProfilingGuide/index.html#metrics-structure}
\showURL{%
\tempurl}


\bibitem[NVIDIA(2022)]%
        {hopper-whitepaper}
\bibfield{author}{\bibinfo{person}{NVIDIA}.} \bibinfo{year}{2022}\natexlab{}.
\newblock \bibinfo{title}{{NVIDIA} {H100} {Tensor} {Core} {GPU} {Architecture}. \url{https://resources.nvidia.com/en-us-tensor-core/gtc22-whitepaper-hopper}}.
\newblock


\bibitem[Paillier(1999)]%
        {paillier}
\bibfield{author}{\bibinfo{person}{P. Paillier}.} \bibinfo{year}{1999}\natexlab{}.
\newblock \showarticletitle{Public-key cryptosystems based on composite degree residuosity classes}. In \bibinfo{booktitle}{\emph{Advances in Cryptology (EUROCRYPT)}}.
\newblock


\bibitem[Piotrowska et~al\mbox{.}(2017)]%
        {203838}
\bibfield{author}{\bibinfo{person}{Ania~M. Piotrowska}, \bibinfo{person}{Jamie Hayes}, \bibinfo{person}{Tariq Elahi}, \bibinfo{person}{Sebastian Meiser}, {and} \bibinfo{person}{George Danezis}.} \bibinfo{year}{2017}\natexlab{}.
\newblock \showarticletitle{The Loopix Anonymity System}. In \bibinfo{booktitle}{\emph{26th USENIX Security Symposium (USENIX Security 17)}}. \bibinfo{publisher}{USENIX Association}, \bibinfo{address}{Vancouver, BC}, \bibinfo{pages}{1199--1216}.
\newblock
\showISBNx{978-1-931971-40-9}
\urldef\tempurl%
\url{https://www.usenix.org/conference/usenixsecurity17/technical-sessions/presentation/piotrowska}
\showURL{%
\tempurl}


\bibitem[Popa et~al\mbox{.}(2011)]%
        {cryptdb}
\bibfield{author}{\bibinfo{person}{Raluca~Ada Popa}, \bibinfo{person}{Catherine M.~S. Redfield}, \bibinfo{person}{Nickolai Zeldovich}, {and} \bibinfo{person}{Hari Balakrishnan}.} \bibinfo{year}{2011}\natexlab{}.
\newblock \showarticletitle{CryptDB: Protecting Confidentiality with Encrypted Query Processing}. In \bibinfo{booktitle}{\emph{In Proceedings of the 23rd ACM Symposium on Operating Systems Principles (SOSP)}}.
\newblock


\bibitem[Research(2024)]%
        {PPML-msft}
\bibfield{author}{\bibinfo{person}{Microsoft Research}.} \bibinfo{year}{2024}\natexlab{}.
\newblock \bibinfo{howpublished}{\url{https://www.microsoft.com/en-us/research/group/privacy-preserving-machine-learning-innovation/}}.
\newblock


\bibitem[Rezapour~Siahgourabi(2020)]%
        {rezapour2020autocpa}
\bibfield{author}{\bibinfo{person}{Zahra Rezapour~Siahgourabi}.} \bibinfo{year}{2020}\natexlab{}.
\newblock \emph{\bibinfo{title}{AutoCPA: Automatic Continuous Profiling and Analysis}}.
\newblock \bibinfo{thesistype}{Master's\ thesis}. \bibinfo{school}{University of Waterloo}.
\newblock


\bibitem[Safi(2021)]%
        {safi2021distributive}
\bibfield{author}{\bibinfo{person}{Miraqa Safi}.} \bibinfo{year}{2021}\natexlab{}.
\newblock \showarticletitle{Distributive continuous profiling for IoT devices}.
\newblock  (\bibinfo{year}{2021}).
\newblock


\bibitem[Sattler et~al\mbox{.}(2022)]%
        {sattler2022icloud}
\bibfield{author}{\bibinfo{person}{Patrick Sattler}, \bibinfo{person}{Juliane Aulbach}, \bibinfo{person}{Johannes Zirngibl}, {and} \bibinfo{person}{Georg Carle}.} \bibinfo{year}{2022}\natexlab{}.
\newblock \showarticletitle{Towards a Tectonic Traffic Shift? Investigating Apple's New Relay Network}.
\newblock \bibinfo{journal}{\emph{arXiv preprint arXiv:2207.02112}} (\bibinfo{year}{2022}).
\newblock
\urldef\tempurl%
\url{https://arxiv.org/abs/2207.02112}
\showURL{%
\tempurl}


\bibitem[Seemakhupt et~al\mbox{.}(2023)]%
        {seemakhupt2023cloud}
\bibfield{author}{\bibinfo{person}{Korakit Seemakhupt}, \bibinfo{person}{Brent~E Stephens}, \bibinfo{person}{Samira Khan}, \bibinfo{person}{Sihang Liu}, \bibinfo{person}{Hassan Wassel}, \bibinfo{person}{Soheil~Hassas Yeganeh}, \bibinfo{person}{Alex~C Snoeren}, \bibinfo{person}{Arvind Krishnamurthy}, \bibinfo{person}{David~E Culler}, {and} \bibinfo{person}{Henry~M Levy}.} \bibinfo{year}{2023}\natexlab{}.
\newblock \showarticletitle{A Cloud-Scale Characterization of Remote Procedure Calls}. In \bibinfo{booktitle}{\emph{Proceedings of the 29th Symposium on Operating Systems Principles}}. \bibinfo{pages}{498--514}.
\newblock


\bibitem[Shen et~al\mbox{.}(2023)]%
        {shen2023propeller}
\bibfield{author}{\bibinfo{person}{Han Shen}, \bibinfo{person}{Krzysztof Pszeniczny}, \bibinfo{person}{Rahman Lavaee}, \bibinfo{person}{Snehasish Kumar}, \bibinfo{person}{Sriraman Tallam}, {and} \bibinfo{person}{Xinliang~David Li}.} \bibinfo{year}{2023}\natexlab{}.
\newblock \showarticletitle{Propeller: A profile guided, relinking optimizer for warehouse-scale applications}. In \bibinfo{booktitle}{\emph{Proceedings of the 28th ACM International Conference on Architectural Support for Programming Languages and Operating Systems, Volume 2}}. \bibinfo{pages}{617--631}.
\newblock


\bibitem[Sherwood et~al\mbox{.}(2001)]%
        {sherwood2001basic}
\bibfield{author}{\bibinfo{person}{Timothy Sherwood}, \bibinfo{person}{Erez Perelman}, {and} \bibinfo{person}{Brad Calder}.} \bibinfo{year}{2001}\natexlab{}.
\newblock \showarticletitle{Basic block distribution analysis to find periodic behavior and simulation points in applications}. In \bibinfo{booktitle}{\emph{Proceedings 2001 International Conference on Parallel Architectures and Compilation Techniques}}. IEEE, \bibinfo{pages}{3--14}.
\newblock


\bibitem[Sherwood et~al\mbox{.}(2002)]%
        {sherwood2002automatically}
\bibfield{author}{\bibinfo{person}{Timothy Sherwood}, \bibinfo{person}{Erez Perelman}, \bibinfo{person}{Greg Hamerly}, {and} \bibinfo{person}{Brad Calder}.} \bibinfo{year}{2002}\natexlab{}.
\newblock \showarticletitle{Automatically characterizing large scale program behavior}.
\newblock \bibinfo{journal}{\emph{ACM SIGPLAN Notices}} \bibinfo{volume}{37}, \bibinfo{number}{10} (\bibinfo{year}{2002}), \bibinfo{pages}{45--57}.
\newblock


\bibitem[Sherwood et~al\mbox{.}(2003a)]%
        {sherwood2003discovering}
\bibfield{author}{\bibinfo{person}{Timothy Sherwood}, \bibinfo{person}{Erez Perelman}, \bibinfo{person}{Greg Hamerly}, \bibinfo{person}{Suleyman Sair}, {and} \bibinfo{person}{Brad Calder}.} \bibinfo{year}{2003}\natexlab{a}.
\newblock \showarticletitle{Discovering and exploiting program phases}.
\newblock \bibinfo{journal}{\emph{IEEE micro}} \bibinfo{volume}{23}, \bibinfo{number}{6} (\bibinfo{year}{2003}), \bibinfo{pages}{84--93}.
\newblock


\bibitem[Sherwood et~al\mbox{.}(2003b)]%
        {sherwood2003phase}
\bibfield{author}{\bibinfo{person}{Timothy Sherwood}, \bibinfo{person}{Suleyman Sair}, {and} \bibinfo{person}{Brad Calder}.} \bibinfo{year}{2003}\natexlab{b}.
\newblock \showarticletitle{Phase tracking and prediction}.
\newblock \bibinfo{journal}{\emph{ACM SIGARCH Computer Architecture News}} \bibinfo{volume}{31}, \bibinfo{number}{2} (\bibinfo{year}{2003}), \bibinfo{pages}{336--349}.
\newblock


\bibitem[Signals({[n.\,d.]})]%
        {parca}
\bibfield{author}{\bibinfo{person}{Polar Signals}.} \bibinfo{year}{[n.\,d.]}\natexlab{}.
\newblock \bibinfo{title}{parca. \url{https://github.com/parca-dev/parca}}.
\newblock


\bibitem[Splunk({[n.\,d.]})]%
        {splunk}
\bibfield{author}{\bibinfo{person}{Splunk}.} \bibinfo{year}{[n.\,d.]}\natexlab{}.
\newblock \bibinfo{title}{Splunk AlwaysOn Profiling. \url{https://docs.splunk.com/observability/en/apm/profiling/intro-profiling.html}}.
\newblock


\bibitem[Sriraman and Dhanotia(2020)]%
        {sriraman2020accelerometer}
\bibfield{author}{\bibinfo{person}{Akshitha Sriraman} {and} \bibinfo{person}{Abhishek Dhanotia}.} \bibinfo{year}{2020}\natexlab{}.
\newblock \showarticletitle{Accelerometer: Understanding acceleration opportunities for data center overheads at hyperscale}. In \bibinfo{booktitle}{\emph{Proceedings of the Twenty-Fifth International Conference on Architectural Support for Programming Languages and Operating Systems}}. \bibinfo{pages}{733--750}.
\newblock


\bibitem[Su et~al\mbox{.}(2024)]%
        {10255554}
\bibfield{author}{\bibinfo{person}{Fei Su}, \bibinfo{person}{Robert Kwasnick}, \bibinfo{person}{John Holm}, \bibinfo{person}{William Penner}, \bibinfo{person}{Hermann Gartler}, \bibinfo{person}{Josh Boelter}, \bibinfo{person}{Yufei Zhou}, \bibinfo{person}{Bijan Arbab}, {and} \bibinfo{person}{Michael Rothberg}.} \bibinfo{year}{2024}\natexlab{}.
\newblock \showarticletitle{Product Health Insights Using Telemetry}.
\newblock \bibinfo{journal}{\emph{IEEE Design \& Test}} \bibinfo{volume}{41}, \bibinfo{number}{4} (\bibinfo{year}{2024}), \bibinfo{pages}{56--64}.
\newblock
\href{https://doi.org/10.1109/MDAT.2023.3317337}{doi:\nolinkurl{10.1109/MDAT.2023.3317337}}


\bibitem[Sun et~al\mbox{.}(2023)]%
        {sun2023omniwindow}
\bibfield{author}{\bibinfo{person}{Haifeng Sun}, \bibinfo{person}{Jiaheng Li}, \bibinfo{person}{Jintao He}, \bibinfo{person}{Jie Gui}, {and} \bibinfo{person}{Qun Huang}.} \bibinfo{year}{2023}\natexlab{}.
\newblock \showarticletitle{OmniWindow: A general and efficient window mechanism framework for network telemetry}. In \bibinfo{booktitle}{\emph{Proceedings of the ACM SIGCOMM 2023 Conference}}. \bibinfo{pages}{867--880}.
\newblock


\bibitem[T.(2024)]%
        {cointribune2024bitcoin}
\bibfield{author}{\bibinfo{person}{Nicolas T.}} \bibinfo{year}{2024}\natexlab{}.
\newblock \bibinfo{title}{Bitcoin - Putting its colossal computing power into perspective}.
\newblock
\shownote{Accessed: April 10, 2025}.
\newblock
\urldef\tempurl%
\url{https://www.cointribune.com/en/bitcoin-putting-its-colossal-computing-power-into-perspective/}
\showURL{%
\tempurl}


\bibitem[Tesla({[n.\,d.]})]%
        {telsa-telemetry}
\bibfield{author}{\bibinfo{person}{Tesla}.} \bibinfo{year}{[n.\,d.]}\natexlab{}.
\newblock \bibinfo{title}{Tesla Fleet Telemetry. \url{https://developer.tesla.com/docs/fleet-api/fleet-telemetry}}.
\newblock


\bibitem[Tian et~al\mbox{.}(2022)]%
        {poisoning}
\bibfield{author}{\bibinfo{person}{Zhiyi Tian}, \bibinfo{person}{Lei Cui}, \bibinfo{person}{Jie Liang}, {and} \bibinfo{person}{Shui Yu}.} \bibinfo{year}{2022}\natexlab{}.
\newblock \showarticletitle{A Comprehensive Survey on Poisoning Attacks and Countermeasures in Machine Learning}.
\newblock \bibinfo{journal}{\emph{Comput. Surveys}} (\bibinfo{year}{2022}).
\newblock


\bibitem[tormetrics({[n.\,d.]})]%
        {tormetrics}
tormetrics \bibinfo{year}{[n.\,d.]}\natexlab{}.
\newblock \bibinfo{title}{Tor Metrics. \url{https://metrics.torproject.org/bandwidth.html}}.
\newblock


\bibitem[Uta et~al\mbox{.}(2020)]%
        {uta2020big}
\bibfield{author}{\bibinfo{person}{Alexandru Uta}, \bibinfo{person}{Alexandru Custura}, \bibinfo{person}{Dmitry Duplyakin}, \bibinfo{person}{Ivo Jimenez}, \bibinfo{person}{Jan Rellermeyer}, \bibinfo{person}{Carlos Maltzahn}, \bibinfo{person}{Robert Ricci}, {and} \bibinfo{person}{Alexandru Iosup}.} \bibinfo{year}{2020}\natexlab{}.
\newblock \showarticletitle{Is big data performance reproducible in modern cloud networks?}. In \bibinfo{booktitle}{\emph{17th USENIX symposium on networked systems design and implementation (NSDI 20)}}. \bibinfo{pages}{513--527}.
\newblock


\bibitem[Weng et~al\mbox{.}(2023)]%
        {weng2023effective}
\bibfield{author}{\bibinfo{person}{Lingmei Weng}, \bibinfo{person}{Yigong Hu}, \bibinfo{person}{Peng Huang}, \bibinfo{person}{Jason Nieh}, {and} \bibinfo{person}{Junfeng Yang}.} \bibinfo{year}{2023}\natexlab{}.
\newblock \showarticletitle{Effective Performance Issue Diagnosis with Value-Assisted Cost Profiling}. In \bibinfo{booktitle}{\emph{Proceedings of the Eighteenth European Conference on Computer Systems}}. \bibinfo{pages}{1--17}.
\newblock


\bibitem[Xu et~al\mbox{.}(2021)]%
        {xu2021PPML}
\bibfield{author}{\bibinfo{person}{Runhua Xu}, \bibinfo{person}{Nathalie Baracaldo}, {and} \bibinfo{person}{James Joshi}.} \bibinfo{year}{2021}\natexlab{}.
\newblock \bibinfo{title}{Privacy-Preserving Machine Learning: Methods, Challenges and Directions}.
\newblock
\showeprint[arxiv]{2108.04417}~[cs.LG]
\urldef\tempurl%
\url{https://arxiv.org/abs/2108.04417}
\showURL{%
\tempurl}


\bibitem[Zhao et~al\mbox{.}(2021)]%
        {265015}
\bibfield{author}{\bibinfo{person}{Yikai Zhao}, \bibinfo{person}{Kaicheng Yang}, \bibinfo{person}{Zirui Liu}, \bibinfo{person}{Tong Yang}, \bibinfo{person}{Li Chen}, \bibinfo{person}{Shiyi Liu}, \bibinfo{person}{Naiqian Zheng}, \bibinfo{person}{Ruixin Wang}, \bibinfo{person}{Hanbo Wu}, \bibinfo{person}{Yi Wang}, {and} \bibinfo{person}{Nicholas Zhang}.} \bibinfo{year}{2021}\natexlab{}.
\newblock \showarticletitle{{LightGuardian}: A {Full-Visibility}, Lightweight, In-band Telemetry System Using Sketchlets}. In \bibinfo{booktitle}{\emph{18th USENIX Symposium on Networked Systems Design and Implementation (NSDI 21)}}. \bibinfo{publisher}{USENIX Association}, \bibinfo{pages}{991--1010}.
\newblock
\showISBNx{978-1-939133-21-2}
\urldef\tempurl%
\url{https://www.usenix.org/conference/nsdi21/presentation/zhao}
\showURL{%
\tempurl}


\bibitem[Úlfar Erlingsson et~al\mbox{.}(2020)]%
        {shuffleDP}
\bibfield{author}{\bibinfo{person}{Úlfar Erlingsson}, \bibinfo{person}{Vitaly Feldman}, \bibinfo{person}{Ilya Mironov}, \bibinfo{person}{Ananth Raghunathan}, \bibinfo{person}{Kunal Talwar}, {and} \bibinfo{person}{Abhradeep Thakurta}.} \bibinfo{year}{2020}\natexlab{}.
\newblock \bibinfo{title}{Amplification by Shuffling: From Local to Central Differential Privacy via Anonymity}.
\newblock
\showeprint[arxiv]{1811.12469}~[cs.LG]
\urldef\tempurl%
\url{https://arxiv.org/abs/1811.12469}
\showURL{%
\tempurl}


\end{thebibliography}

\appendix
\section{Formal Protocol Definition}\label{app:protocol}

This section provides a formal definition of the \sysname data submission protocol.

\subsection{Protocol: \sysname Data Submission}

    \textbf{Parties:} A set of Clients $\mathcal{C}$, an Aggregation Server AS, and a Designer Server DS.

    \textbf{Setup:}
    \begin{enumerate}
        \item The DS generates a Paillier key pair $(pk, sk)$, where $pk=(g, n)$, and makes the public key $pk$ available to all clients (e.g., via a public website with a certificate).
        \item All parties agree on a set of cryptographic hash functions $H_1$ (e.g., SHA-256) and a MinHash function family $\mathcal{H}$.
    \end{enumerate}

    \textbf{Protocol Execution (for a single client $C \in \mathcal{C}$):}
    \begin{enumerate}
        \item \textbf{Client-side Snippet Generation:}
        \begin{itemize}
            \item $C$ observes a kernel sequence $S = (k_1, \dots, k_L)$.
            \item $C$ computes the MinHash signature: $M_S = \text{MinHash}_{\mathcal{H}}(S)$.
            \item $C$ computes the Snippet Hash: $h_S = H_1(M_S)$.
        \end{itemize}

        \item \textbf{Client-side Data Aggregation:}
        \begin{itemize}
            \item $C$ collects performance counter samples for snippet $S$ and aggregates them into a partial histogram $P = \{b_1, b_2, \dots, b_{128}\}$.
        \end{itemize}

        \item \textbf{Client-side Encryption and Transmission:}
        \begin{itemize}
            \item For each bin value $b_i \in P$, $C$ computes the ciphertext $c_i = \text{Enc}_{pk}(b_i)$. Let the encrypted histogram be $P_{\text{enc}} = \{c_1, \dots, c_{128}\}$.
            \item $C$ establishes a new, anonymous communication channel to AS (e.g., a new Tor circuit).
            \item $C$ sends the tuple $(h_S, M_S, P_{\text{enc}})$ to AS.
        \end{itemize}

        \item \textbf{Server-side Aggregation (AS):}
        \begin{itemize}
            \item Upon receiving $(h_S, M_S, P_{\text{enc}})$, AS first uses $h_S$ and $M_S$ to identify the corresponding Canonical Snippet's Aggregated Snippet Histogram (ASH).
            \item For $i=1$ to $128$, AS updates the corresponding bin $C_i$ in the ASH: $C_i \leftarrow C_i \cdot c_i \pmod{n^2}$.
        \end{itemize}

        \item \textbf{Server-side Reporting (AS to DS):}
        \begin{itemize}
            \item Periodically, AS sends the fully updated, encrypted ASHs to DS.
        \end{itemize}

        \item \textbf{Designer-side Decryption (DS):}
        \begin{itemize}
            \item DS receives an encrypted ASH, $\{C_1, \dots, C_{128}\}$.
            \item For each bin, DS decrypts the value: $B_i = \text{Dec}_{sk}(C_i)$.
        \end{itemize}
    \end{enumerate}

\subsection{Security Proof Sketch}\label{app:proof}

This section provides a sketch of the arguments for the security and privacy guarantees of the \sysname protocol.

\subsection*{Threat Model}
We assume an honest-but-curious (or semi-honest) Aggregation Server (AS) and Designer Server (DS). Both parties follow the protocol correctly but may attempt to infer additional information from the data they observe. We assume the AS and DS do not collude.

\subsection*{Security Guarantees}
\begin{enumerate}
    \item \textbf{Histogram Confidentiality:} This property rests on the \textbf{semantic security} of the Paillier cryptosystem. An adversary (the AS) only observes Paillier ciphertexts. The ability of the AS to distinguish between the encryption of two different histograms would imply an ability to break the semantic security of Paillier, which is a standard, well-accepted hardness assumption. Therefore, the AS learns nothing about the content of the partial or aggregated histograms.

    \item \textbf{User Anonymity:} This guarantee relies on the \textbf{anony-mity properties of the underlying network (e.g., Tor)}. Assuming the network prevents the AS from learning the user's true IP address, the AS cannot link a submission to a real-world identity. Furthermore, because histograms are encrypted (by Guarantee 1) and a new anonymous circuit is used for each submission, the AS cannot link multiple submissions from the same user over time (unlinkability).

    \item \textbf{Application Confidentiality:} This rests on the \textbf{pre-image resistance} of the cryptographic hash functions used for the Snippet Hash and MinHash. An adversary seeking to learn a private kernel sequence would need to perform a brute-force or dictionary attack to invert the MinHash. The infeasibility of such an attack relies on three layers of defense:
    \begin{itemize}
        \item \textbf{Large Search Space:} The attack's complexity is defined by the size of the kernel "alphabet" ($N$) and the n-gram length ($n$). With $N \approx 10^4$ public kernels and $n=8$, the search space is on the order of $N^n = (10^4)^8 = 10^{32}$, making a brute-force search computationally impossible.
        \item \textbf{Incomplete Dictionary:} The existence of private, proprietary kernels means that an adversary's dictionary of public kernels is inherently incomplete, rendering a dictionary attack ineffective against applications that use them.
        \item \textbf{Kernel Name Obfuscation:} As discussed in \S\ref{sec:privacy}, the use of a per-application salt during compilation further hardens this defense, ensuring that even known public kernels will have unique hashes within a private application.
    \end{itemize}
    Therefore, the security reduces to the assumed pre-image resistance of the hash functions against an adversary with an incomplete dictionary of possible inputs.
\end{enumerate}

\begin{table*}[h]
\caption{Some Low Level Counters on A100}
\label{tab:counterlist}
\begin{tabular}{ll}
\toprule
\textbf{Metric} & \textbf{Description} \\
\midrule
\texttt{sm\_\_cycles\_active} & \# of cycles with at least one warp in flight \\
\texttt{sm\_\_pipe\_alu\_cycles\_active} & \# of cycles where alu pipe was active \\
\texttt{sm\_\_pipe\_fma\_cycles\_active} & \# of cycles where fma pipe was active \\
\texttt{sm\_\_pipe\_fp64\_cycles\_active} & \# of cycles where fp64 pipe was active \\
\texttt{sm\_\_pipe\_tensor\_op\_hmma\_cycles\_active} & \# of cycles where tensor pipe was active (HMMA/HFMA2.MMA ops) \\
\texttt{sm\_\_pipe\_tensor\_op\_imma\_cycles\_active} & \# of cycles where tensor pipe was active (IMMA/BMMA ops) \\
\texttt{smsp\_\_cycles\_active} & \# of cycles with at least one warp in flight     \\
\texttt{dram\_\_throughput} & DRAM throughput \\
\texttt{smsp\_\_warps\_launched} & \# of warps launched \\
\texttt{smsp\_\_warps\_issue\_stalled\_membar} & cumulative \# of warps waiting on a memory barrier \\
\texttt{sm\_\_sass\_inst\_executed\_op\_global} & \# of warp instructions executed: LDG, STG, LD, ST, ATOM, AT\\
\texttt{l1tex\_\_data\_bank\_conflicts\_pipe\_lsu} & \# of data bank conflicts generated by LSU pipe \\
\texttt{l1tex\_\_data\_bank\_reads} & \# of data bank reads \\
\texttt{l1tex\_\_data\_bank\_writes} & \# of data bank writes \\
\texttt{l1tex\_\_m\_l1tex2xbar\_write\_bytes} & \# of bytes written to L2 \\
\texttt{l1tex\_\_m\_xbar2l1tex\_read\_bytes} & \# of bytes read from L2 into L1TEX M-Stage \\
\bottomrule
\end{tabular}
\end{table*}
\section{List of Counters}\label{app:counters}
On A100, there are 100s of low-level counters, which can be obtained with ncu --metrics. In Table~\ref{tab:counterlist} are a list of some meaningful ones often used in performance characterization and to understand hardware $\times$ workload interactions.

\section{Chip Designers Using Penrose}\label{app:designers}

\subsection{Histogram Representation}\label{sec:representation}

\sysname emits both one- and two-dimensional histograms. A 1-D histogram of a
single counter reports how much pressure a workload places on that resource,
but cannot express co-occurrence: it records that a high-tensor mode and a
low-DRAM mode each exist without establishing that they are the same samples.
Replacing a run with a surrogate that has identical 1-D marginals but
independent pairing drops
$P(\texttt{tensor\_active} > 0.3 \wedge \texttt{dram\_active} < 0.35)$ from
$0.152$ to $0.019$. The 2-D histogram supplies exactly that missing structure.

It is not, however, a lossless encoding of the trace. The 2-D histogram
preserves which modes exist and their relative time-share, while discarding
their order, their individual durations, and how often execution alternates
between them. The run in Figure~\ref{fig:shipped} contains three prefill
episodes lasting \SI{3.7}{\second}, \SI{0.2}{\second}, and \SI{0.2}{\second};
its histogram records only that $15.2\%$ of active time fell in the compute
mode, and cannot distinguish this from a single uninterrupted
\SI{4.1}{\second} episode. That loss is deliberate: the discarded structure is
precisely the request-timing and burst information that would make a trace
identifying, while the retained structure is what a chip designer needs.

\subsection{Per-Model Joint Occupancy}\label{sec:per-model}

\begin{figure*}[tbp]
    \includegraphics[width=\textwidth]{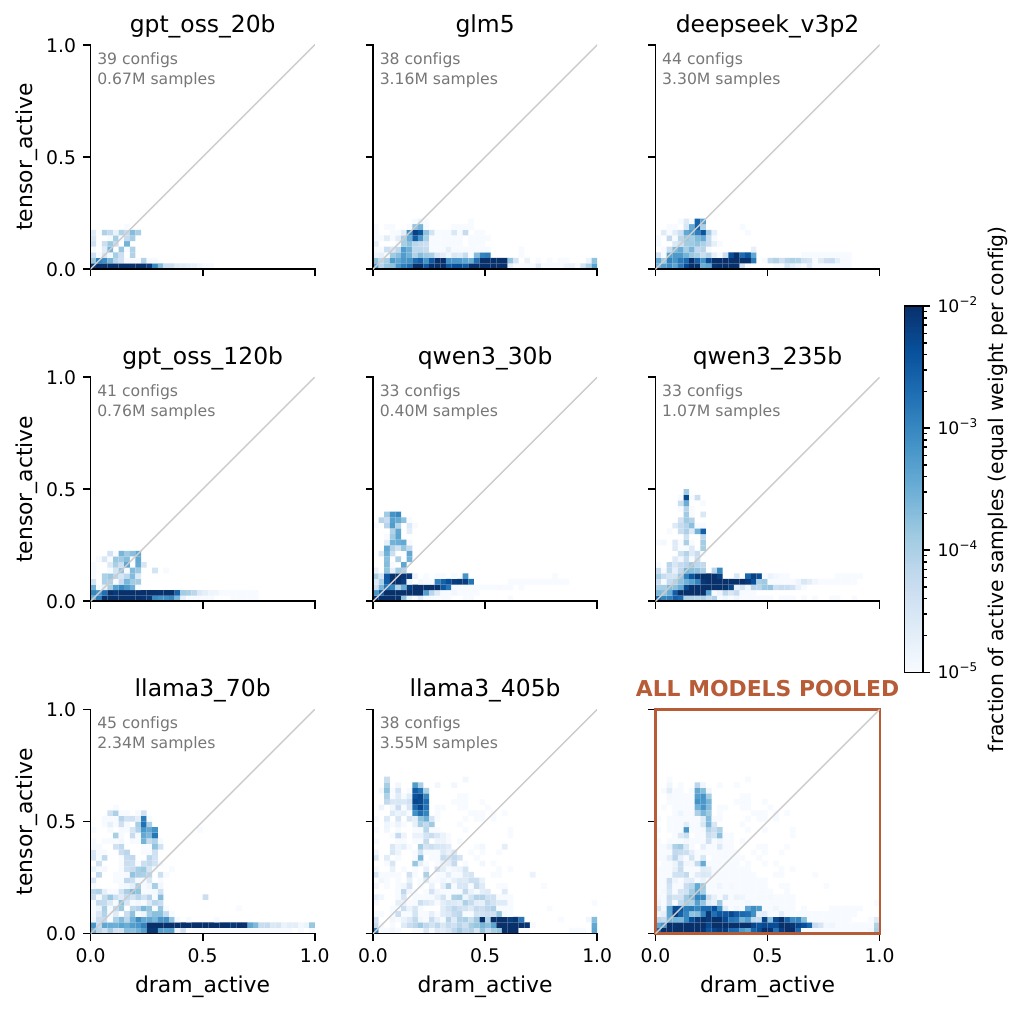}
    \caption{Joint tensor/DRAM occupancy per model, merging every batch size,
    context length, and generation length. Each configuration is weighted
    equally. Panels are ordered by prefill compute peak; the final panel pools
    all models, retaining bimodality but losing the per-model mode positions.}
    \label{fig:permodel}
\end{figure*}

Figure~\ref{fig:permodel} merges every sweep configuration of each model---all
batch sizes, context lengths, and generation lengths---into one joint histogram
per model across all eight GPUs (311 runs, \num{15.2}~million active samples; 10
truncated runs excluded). Each run is normalized before merging so that
configurations contribute equally; otherwise 65536-token runs outweigh
256-token runs by ${\sim}100\times$ and the prefill mode vanishes. Panels are
ordered by prefill compute peak, which rises from $0.11$ (GPT-OSS-20B) to
$0.64$ (Llama-3-405B), while the decode plateau spans $0.06$ to $0.55$---a
$6$--$9\times$ spread across models that no aggregate exposes. Merging across
configurations also widens each decode mode into a horizontal band whose extent
measures the range of bandwidth demand a model exerts across its operating
points (Llama-3-70B spans \texttt{dram\_active} $0.1$--$0.7$). The final panel
pools all models: bimodality survives, but the model-specific mode positions do
not---which is the case for reporting per-model rather than aggregate
histograms.


\section{Generative AI Use Disclosure}
Claude (Opus 5) was utilized to generate sections of this Work, including text, figures, and code.

\end{document}